\documentclass[aps, prx, amssymb, longbibliography, twocolumn, superscriptaddress]{revtex4-2}

\usepackage[german,english]{babel}
\usepackage[utf8]{inputenc}

\usepackage{graphicx}
\usepackage{dcolumn}
\usepackage{bm}
\usepackage{physics}

\usepackage{dcolumn}
\usepackage{subfigure}

\usepackage{tabularx}
\usepackage[normalem]{ulem}
\usepackage[usenames, dvipsnames]{color}

\usepackage{placeins}

\usepackage{listings}
\usepackage{xcolor}
\usepackage{float}

\usepackage{algpseudocode}

\usepackage{colortbl}
\usepackage{diagbox}
\usepackage[version=4]{mhchem}

\usepackage{theorem}
\newtheorem{theorem}{Theorem}

\newtheorem{observation}{Observation}

\newenvironment{proof_theorem}[1]{\noindent \textbf{{Proof of Theorem~{#1}~} }}{\hfill $\blacksquare$}

\definecolor{myrefcolor}{rgb}{0.067,0.5,0.5}

\usepackage[
    breaklinks,
    pdftex,
    colorlinks=true,
    linkcolor=myrefcolor,
    citecolor=myrefcolor,
    urlcolor=myrefcolor
]{hyperref}

\begin{document}

\title{Unveiling quantum phase transitions from traps in variational quantum algorithms}
	
	\author{Chenfeng Cao}
	\affiliation{Phasecraft Ltd.}
    \affiliation{Dahlem Center for Complex Quantum Systems, Freie Universit\"{a}t Berlin}
	
	\author{Filippo Maria Gambetta}
	\affiliation{Phasecraft Ltd.}
	
	\author{Ashley Montanaro}
	\email{ashley@phasecraft.io}
	\affiliation{Phasecraft Ltd.}
	\affiliation{School of Mathematics, University of Bristol}
	
	\author{Raul A. Santos}
	\affiliation{Phasecraft Ltd.}
	
	\date{\today}
	
	\begin{abstract}
		Understanding quantum phase transitions in physical systems is fundamental to characterize their behavior at low temperatures. Achieving this requires both accessing good approximations to the ground state and identifying order parameters to distinguish different phases. Addressing these challenges, our work introduces a hybrid algorithm that combines quantum optimization with classical machine learning. This approach leverages the capability of near-term quantum computers to prepare locally trapped states through finite optimization. Specifically, we apply LASSO for identifying conventional phase transitions and the Transformer model for topological transitions, utilizing these with a sliding window scan of Hamiltonian parameters to learn appropriate order parameters and locate critical points.  We validated the method with numerical simulations and real-hardware experiments on Rigetti's Ankaa 9Q-1 quantum computer. This protocol provides a framework for investigating quantum phase transitions with shallow circuits, offering enhanced efficiency and, in some settings, higher precision-thus contributing to the broader effort to integrate near-term quantum computing and machine learning.

	\end{abstract}
	
	\maketitle

	\section{Introduction}
	\begin{figure}[tbh]
		\centering
		\includegraphics[width=7.7cm]{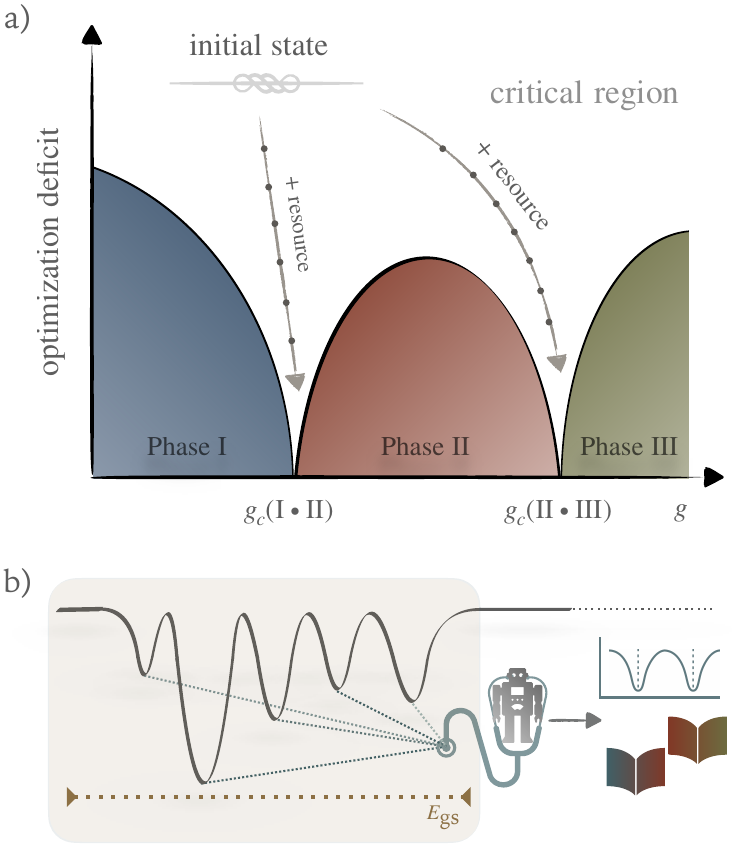}
		\caption{Schematic picture of (a) a finite-optimization phase diagram and (b) learning effective order parameters and phase transitions from local minima states. In (a), as the optimization deficit increases—attributable to either less expressive variational ansätze or inadequate optimization—the system becomes less capable of reaching the ground state, causing a slight shift of the system's phase transition away from the critical point associated with the ground state. In (b), machine learning identifies potential phase transitions by constructing a classical loss landscape. Within this landscape, potential phase transitions are located by valleys.}
		\label{fig:schematic}
	\end{figure}

	Quantum phase diagrams are a fundamental tool to characterize the behavior of quantum systems under varying external conditions, such as temperature and magnetic fields~\cite{subir1999quantum, sachdev2003order, vojta2003quantum, braun2009phase, carr2010understanding, sachdev2011qpt}. A better understanding of quantum phase transitions finds application in the field of materials science, where it may inform the development of novel materials with unique properties. For instance, understanding superconductor–insulator phase transitions is instrumental in the research and development of high-temperature superconductors~\cite{gantmakher2010superconductor, keimer2015from}. Due to the intrinsic complexity of quantum systems, the accurate determination of the ground-state phase diagram is a formidable challenge~\cite{kitaev2002classical, kempe2006the}. Experimentally achieving and maintaining the ground state is also difficult, requiring precise control of conditions and often near-zero temperatures~\cite{georgescu2014quantum, suter2016protecting, tomza2019cold}. Furthermore, the characterization of quantum critical points is often hindered by unknown order parameters, particularly in systems undergoing multiple unconventional phase transitions~\cite{subir1999quantum, vojta2003quantum, kosterlitz2017nobel}.  Traditional approaches, such as Landau's theory, which link order parameters with symmetry changes, often fail for non-trivial topological phase transitions or phases without conventional symmetry breaking~\cite{wen2004quantum, balents2010spin, chiu2016classification}.

	In general, approximating the ground state of a system is known to be hard, even for a quantum computer~\cite{kempe2006the}. Nonetheless, variational quantum approaches offer a practical alternative for creating states that capture correctly the physics of the ground state in some cases~\cite{cerezo2021variational, Stanisic_2022, Cao2023Mitigating}. These methods have become increasingly important in the study of quantum phase transitions~\cite{dreyer2021quantum, an2022learning, meth2022probing, okada2023classically, bosse2023sketching, shi2023locating, lively2024robust, crognaletti2024equivariant}. Particularly relevant and noteworthy are the works of Dreyer \textit{et al.}~\cite{dreyer2021quantum} and Bosse \textit{et al.}~\cite{bosse2023sketching}. In Ref.~\cite{dreyer2021quantum} the authors utilized variational quantum optimization on the transverse-field Ising chain, revealing an intriguing behavior of the order parameter, i.e. the transverse magnetization: it exhibits scaling collapse with respect to the circuit depth exclusively on one side of the phase transition, illustrating distinct behaviors across the critical point. Bosse \textit{et al.}~\cite{bosse2023sketching} explored the utility of the variational quantum eigensolver for learning phase diagrams of quantum systems, focusing on the use of both the change of the energy of the output states during optimization and a variety of traditional order parameters. This study underscores the potential of (low-fidelity) variationally optimized quantum states to provide reliable indicators of phase transitions across a variety of quantum models, including both one-dimensional and two-dimensional spin and fermion systems. However, even with polynomially-sized quantum circuits, classical optimization strategies may be hindered by locally trapped states within the optimization landscape~\cite{anschuetz2022quantum, anschuetz2024unified}. Although these states may effectively reflect the physical properties of various phases, accurately identifying the specific order parameter that defines the phase they represent remains a challenging task.

	At the same time, advancements in machine learning (ML) have opened new ways for analysing quantum many-body problems~\cite{carleo2019machine, vicentini2021machine, dawid2023modern}. Notably, machine learning techniques—especially unsupervised learning—have been extensively employed to detect quantum phase transitions~\cite{wang2016discovering, Nieuwenburg2017A, carrasquilla2017the, huembeli2018identifying, canabarro2019unveiling, lidiak2020unsupervised, Kaming2021unsupervised, huang2022provably, tibaldi2023unsupervised}. Studies such as those by Huang \textit{et al.}~\cite{huang2022provably}, Lewis \textit{et al.}~\cite{lewis2024improved}, and Onorati \textit{et al.}~\cite{onorati2023efficient, onorati2023provably} have further theoretically highlighted the efficacy of these algorithms in learning unknown properties of quantum systems' ground states, significantly within identical phases, even in the absence of exponentially decaying correlations. This capability to distinguish ground-state data from different phases makes machine learning a potent tool for detecting phase transitions by incrementally adjusting the Hamiltonian parameter and modifying the dataset. This philosophy leverages machine learning's differential performance across datasets to accurately identify critical phase changes. Despite these innovations, most studies (e.g., refs.~\cite{Nieuwenburg2017A, canabarro2019unveiling, lidiak2020unsupervised, huang2022provably, tibaldi2023unsupervised}) were focusing on data from exact ground states calculated via methods like exact diagonalization or quantum Monte Carlo simulations, which may limit the analysis of larger or classically intractable systems. Moreover, the determination of the order parameter following training remains ambiguous.

	Inspired by these advancements, this work aims to study the use of locally trapped states to characterize phase transitions, without prior knowledge of the associated order parameters. We employ classical machine learning to distill meaningful patterns from these states, thereby learning effective order parameters. Local traps, including local minima, commonly found in the complex energy landscapes of quantum systems, often hinder the efficiency of optimization algorithms by preventing convergence to the global minimum~\cite{anschuetz2022quantum, chen2023local}. Nevertheless, they can contain important information about the phase of a quantum system. Fig.~\ref{fig:schematic} provides a schematic overview of our approach. When the optimization deficit is relatively small—that is, the locally trapped states have a non-negligible overlap with the ground state—we show that classical machine learning can successfully detect ground-state phase transitions. Indeed, in certain cases, a ground-state phase transition affects the spectrum and properties of the low-lying excited states. Under these circumstances, we expect that the machine learning protocol will remain effective, even when the variationally optimized quantum state possesses minimal overlap with the ground state. As quantum phase transitions are well defined only in the thermodynamic limit, applying our approach to finite-size systems can only provide approximate estimates of the critical points. The finite-size-scaling technique is a well-established method to overcome this issue and to obtain accurate values of critical properties in the thermodynamic limit by extrapolating results obtained within finite-size systems~\cite{fisher1972scaling, newman1999montecarlo}. In this work we show that, under certain assumptions, our algorithm can be employed to implement a variant of this technique in terms of quantum resources, quantified by the depth of the variational quantum circuits, that we term finite-depth extrapolation technique.

	In this study, we deploy a quantum optimization-machine learning hybrid algorithm to detect both traditional and topological phase transitions across various quantum systems, including the 1D and 2D transverse-field Ising models (TFIMs) and the extended Su-Schrieffer-Heeger (eSSH) model. Our methodology incorporates linear and deep learning machine learning techniques, namely LASSO~\cite{tibshirani1996regression} and Transformer~\cite{vaswani2017attention}. In quantum information, LASSO has been employed for tasks like predicting ground state properties~\cite{lewis2024improved} and for learning unknown observables~\cite{molteni2024exponential}. Here, we apply it to extract effective order parameters and accurately determine critical boundaries. Transformer models, while previously utilized, for example, to perform time-domain analysis from quantum Monte Carlo data~\cite{Ding2022Rapid} or serving as ans\"atze within Monte Carlo frameworks~\cite{h2024first}, are implemented in our work in a different way and with a distinct purpose: to capture non-trivial phase-transition-related non-local interdependencies within the data arising from locally trapped states from variational quantum circuits. We show that our protocol can identify critical points and discern candidate order parameters associated with the phase transitions. Notably, for the 1D TFIM, our approach outperforms traditional magnetization measurements and, through the application of the finite-depth extrapolation technique, our approach determines the critical point of the model's quantum phase transition with high accuracy despite the maximum fidelity achieved by our shallow variational circuits near this point being only 0.75. In the 2D TFIM, we show that our algorithm accurately predicts phase transitions despite considerable gate noise, with validation provided through noisy experiments on Rigetti's Ankaa 9Q-1 quantum device. Finally, for the eSSH model, we demonstrate that our method effectively learns an order parameter with minimal finite-size effects, yielding more precise critical‑point estimates than those from the finite-size partial-reflection many-body topological invariant.~\cite{meth2022probing}.

    While our demonstrations in some scenarios employ relatively shallow variational circuits and local observables—raising potential concerns about classical simulability in certain regimes~\cite{Napp2022Efficient, Cerezo2024Does}—our approach is not restricted to this setting. In practice, polynomial- or even logarithmic-depth circuits may remain classically intractable for large systems, particularly in the presence of long-range interactions or complex connectivity. Rather than making a formal claim of quantum advantage, our primary goal is to show that reliable phase transition detection remains possible even if the variational circuit does not closely approximate the ground state. We discuss the interplay between our method and potential classical simulability, along with other limitations, in the Discussion. Crucially, although many previous studies have focused on data derived from exact ground states for theoretical and numerical analyses, our methodology demonstrates that reliable phase transition detection can be achieved using variationally optimized local minima states. These states capture the essential physics even at shallow circuit depths, despite their deviation from the exact ground state.

	\section{Results}\label{sec:applications}
	
	In this section, we apply our algorithm, described in Section~\ref{sec:general_framework}, to study quantum phase transitions in various quantum models. We begin with the 1D transverse-field Ising model (TFIM) with periodic boundary conditions. This model, particularly well-suited for applying our finite-depth extrapolation method, allows us to verify the exponential scaling of Eq.~\eqref{eq:depth_scaling}  and to test the algorithm's resilience to shot noise. Subsequently, our focus shifts to the 2D TFIM with open boundaries, explored through both numerical analysis and experiments on Rigetti's Ankaa 9Q-1 quantum computer, where we assess the algorithm’s robustness to moderate levels of gate noise in our tested settings. Finally, we examine the non-integrable extended Su-Schrieffer-Heeger (eSSH) model~\cite{sylvain2019observation, elben2020many}, characterized by its topological phase transitions. We apply the LASSO-based algorithm to the first two models, where phase transitions are identifiable through local order parameters. In contrast, the eSSH model, which lacks straightforward local order parameters, is analysed using a Set Transformer-based approach. This method demonstrates effectiveness in at identifying complex, non-local interdependencies and directly learning non-linear order parameters from the data. For each configuration considered, we initiate the optimization process with a single set of starting values for $\boldsymbol{\zeta}$ and $\boldsymbol{\eta}$, subsequently employing the Broyden–Fletcher–Goldfarb–Shanno algorithm~\cite{broyden1970the, fletcher1970an, goldfarb1970family, shanno1970conditioning} to obtain local minima states. The quantum circuits are simulated using the Pennylane software framework~\cite{bergholm2022pennylane}, with all loss functions being linearly normalized within the range $[0,1]$.

	\subsection{1D Transverse-field Ising Model}\label{Sec:1DTFIM}

	The initial subject of our investigation is the 1D TFIM with periodic boundary conditions, described by the following Hamiltonian:
	\begin{equation}
		\mathcal{H}_{\text{TI-1D}} = -J\sum_j\sigma^z_j\sigma^z_{j+1} - g\sum_j\sigma^x_j.
	\end{equation}
	The first term describes the interaction between neighbouring spins along the chain and the coupling constant $J$ quantifies the strength of this interaction. The second term represents the influence of a transverse magnetic field applied perpendicular to the direction of spin-spin interaction. The competition between these two terms leads to a quantum phase transition at the critical point $g = J$~\cite{subir1999quantum, sachdev2011qpt}. For $g<J$, the spin-spin interaction dominates, resulting in a phase where spins are ordered. For $g>J$,  the transverse field dominates, leading to a disordered phase where spins align with the field. Usually, this transition is characterized by measuring the magnetization in the $x$-direction, $m_x = 1/n\sum_{j=1}^n\sigma^x_j$, where $n$ is the number of spins in the system. In what follows, the coupling strength $J$ is set to be 1.
	
	Our approach employs the Hamiltonian variational ansatz, as detailed in Refs.~\cite{wecker2015progress, wiersema2020exploring}, parameterized by $\boldsymbol{\beta}, \boldsymbol{\gamma}$ (see Section~\ref{subsec:algorithm}):
	\begin{equation}
		U_{\text{TI-1D}}(\boldsymbol{\beta}, \boldsymbol{\gamma})=\prod_{j=1}^p \exp \left(-i \frac{\beta_j}{2} \mathcal{H}_x\right) \exp \left(-i \frac{\gamma_j}{2} \mathcal{H}_{z z}\right) ,
	\end{equation}
	where $\mathcal{H}_{zz}$ and $\mathcal{H}_x$ correspond to the sum of $\sigma^z_j\sigma^z_{j+1}$ and $\sigma^x_j$, respectively. As described in section~\ref{subsec:algorithm}, $\boldsymbol{\beta}, \boldsymbol{\gamma}$ are represented by a Fourier series, with Fourier coefficients $\tilde{\boldsymbol{\beta}}$ and $\tilde{\boldsymbol{\gamma}}$ expressed as polynomials of $g$ up to the fourth order, 
 	\begin{equation}
		\tilde{\gamma}_k = \sum_{j=0}^4 \zeta_{j, k} g^j, \quad \tilde{\beta}_k = \sum_{j=0}^4 \eta_{j, k} g^j.
	\end{equation}
    The vectors $\boldsymbol{\zeta}$ and $\boldsymbol{\eta}$ are then optimized to minimize the energy function sum
    \begin{equation}
		\mathcal{E}_{\text{opt}} = \sum_{g \in \mathcal{G}_{\text{opt}}} \operatorname{Tr}\left( \mathcal{H}_{\text{TI-1D}}(g) \rho(\boldsymbol{\zeta}, \boldsymbol{\eta}; g)\right),
	\end{equation}
	where $\rho(\boldsymbol{\zeta}, \boldsymbol{\eta}; g)$ are the optimized states and $\mathcal{G}_{\mathrm{opt}} = \{g_\mathrm{min}, g_\mathrm{min} + \delta g, ..., g_\mathrm{max}\}$, with $\delta g$ the optimization sampling resolution, is the optimization grid. The global optimization approach we adopt here ensures better stability and allows us to generate optimized states for any value of $g$. This is particularly useful in training the machine learning algorithm for the detection the system's quantum phase transitions, where we prepare optimized states $\rho(\boldsymbol{\zeta}, \boldsymbol{\eta}; g)$ for all the values of $g$ in the detection grid $\mathcal{G}_{\mathrm{det}} = \{\hat{g}_\mathrm{min}, \hat{g}_\mathrm{min} + \delta \hat{g}, ..., \hat{g}_\mathrm{max}\}$, with $\delta \hat{g}$ the detection sampling resolution. See section~\ref{subsec:algorithm} for more details.
	
	For the optimization process of the 1D TFIM, we choose $g_{\text{min}} = 0$ to $g_{\text{max}} = 2$, with a sampling resolution of $\delta g = 0.1$. The detection grid spans from $\hat{g}_{\text{min}} = 0.57$ to $\hat{g}_{\text{max}} = 1.13$, with a finer sampling resolution of $\delta \hat{g} = 0.001$. As long as $n \geq 2p + 2$, the optimized local observable expectation values are ensured to remain consistent for infinitely large $n$. Beginning with $p = 1$ and $n = 4$, we sample initial values for $\boldsymbol{\zeta}$ and $\boldsymbol{\eta}$ from a narrow range $(-10^{-3}, 10^{-3})$, then we iteratively adjust their values to minimize the energy sum over $\mathcal{G}_{\text{opt}}$. We found that different samples of the initial values of $\boldsymbol{\zeta}$ and $\boldsymbol{\eta}$  within this near-zero range do not lead to significant differences in the outcomes. Hence, we present results based on a singular random sample in subsequent analyses.
	
	Upon convergence, we prepare states for each $g$ in $\mathcal{G}_{\text{det}}$, measuring expectation values for all Pauli terms with weight smaller than $3$:
	\begin{equation}
		\begin{aligned}
			\mathcal{P}_f=\big\{ & \sigma_j^x, \sigma_j^y, \sigma_j^z \\
			& \sigma_i^x \sigma_j^x, \sigma_i^y \sigma_j^y, \sigma_i^z \sigma_j^z \\
			& \sigma_i^x \sigma_j^y, \sigma_i^y \sigma_j^z, \sigma_i^z \sigma_j^x\big\}_{i \neq j} .
		\end{aligned}
	\end{equation}
	Then we employ a sliding window with window size $w = 30$ and LASSO regression with an adaptive regularization parameter $\lambda$ to locate the phase transition critical point. In practical scenarios,  one might use the technique of classical shadows to obtain such local observable expectations simultaneously~\cite{huang2020predicting}. 

	\begin{figure}[tbh]
		\centering
		\includegraphics[width=8cm]{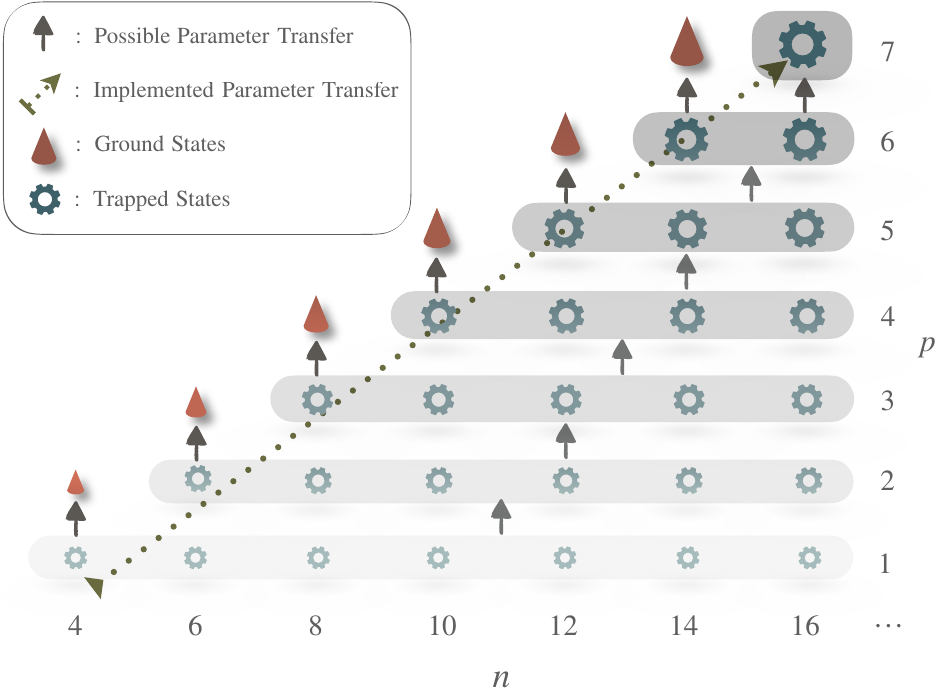}
		\caption{Schematic depiction of the recursive optimization process, demonstrating the incremental increase in circuit depth $p$ and system size $n$, and potential pathways for parameter transfer optimization. Although setting $p = n/2$ allows for the exact preparation of ground states as explored in \cite{wiersema2020exploring}, our study focuses on the dynamics of locally trapped states under conditions where $n \geq 2p + 2$, thus examining the effects of finite circuit depth. Importantly, for any given $p$, systems with $n \geq 2p + 2$ can share the same optimized parameters. In our simulations, we specifically adopt $n = 2p + 2$ for optimization, as highlighted by the dotted dark green arrow, to maintain a consistent approach across different system sizes.}
		\label{fig:recursive_optimization}
	\end{figure}

	In the process of adapting and utilizing optimized parameters for further analysis, we simultaneously increase both $p$ and $n$, specifically, 
	\begin{equation}
		p \rightarrow p + 1, n \rightarrow n + 2,
	\end{equation}
	thereby extending $\Tilde{\boldsymbol{\beta}}$ and $\Tilde{\boldsymbol{\gamma}}$ with zero initial values to accommodate the expanded circuit and system size. This procedure is iteratively executed in numerical simulations, enabling the prediction of critical points $\Tilde{g}_c$ for various values of $p$. The recursive optimization strategy is depicted in Fig.~\ref{fig:recursive_optimization}. \begin{figure}[!h]
		\centering
		\includegraphics[width=7.83cm]{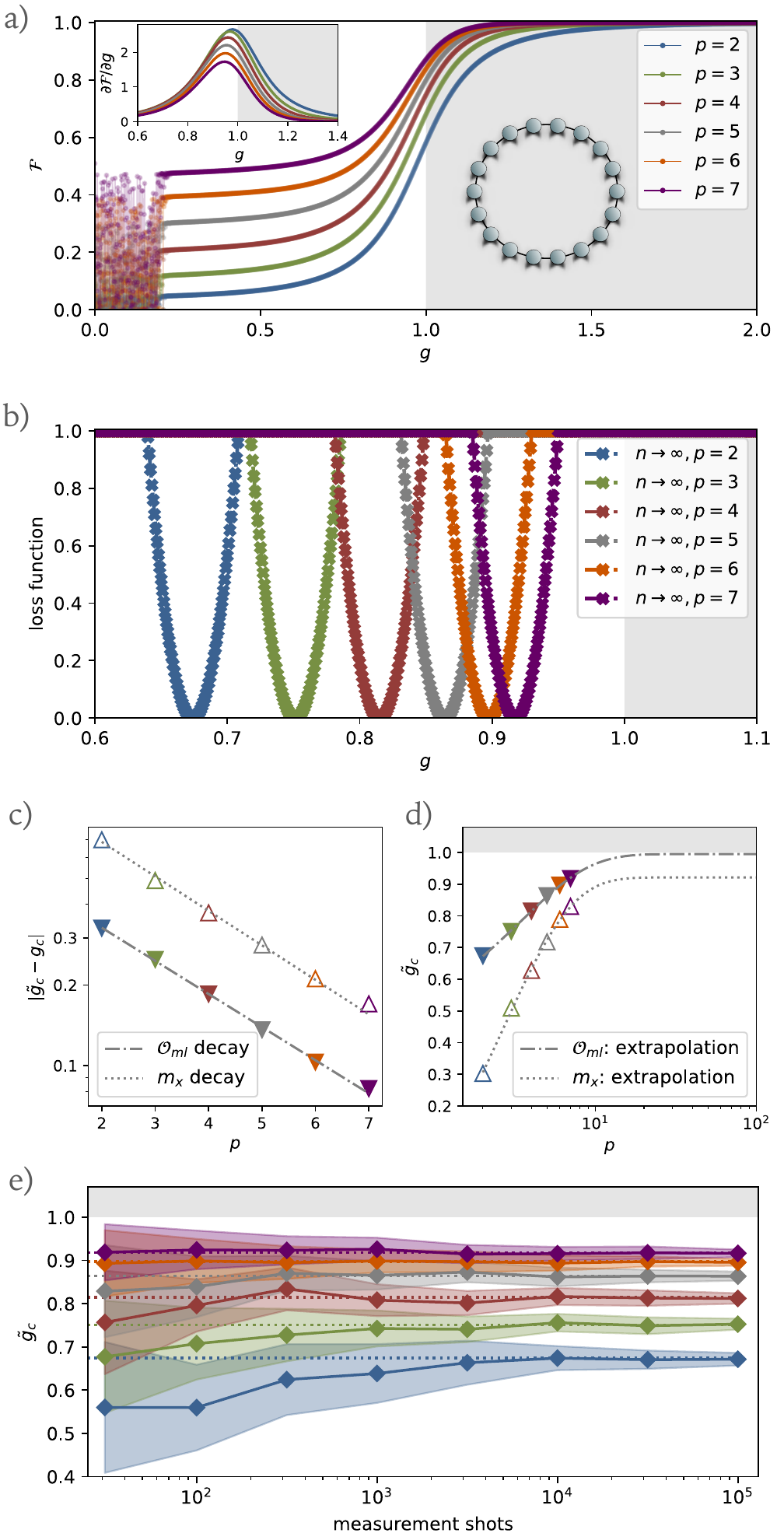}
		\caption{Results of phase transition detection for the 1D transverse-field Ising model. (a) The fidelity between optimized states and the true ground states, evaluated across varying $p$ and $g$  values, with the system size fixed at $n=18$. The inset shows the fidelity derivatives. (b) The normalized loss landscape generated by LASSO regression, with the minimum point indicating the predicted phase transition critical point $\Tilde{g}_c(p)$. (c) The exponential decay trend of the difference between $\Tilde{g}_c$ and the theoretical value $g_c = 1$ as a function $p$. Hollow triangles represent values derived from $x$-magnetization and solid triangles denote values from our loss landscape. (d) Finite-depth exponential extrapolation to estimate the critical point in the thermodynamic limit. The extrapolated values obtained for $x$-magnetization and our loss landscape are $0.921\pm 0.004$ and $0.995\pm0.002$, respectively. (e) The average and standard deviation (represented by the shaded areas) of the predicted critical point $\Tilde{g}_c$ as a function of the number of shots. In all panels, colours correspond to the different values of $p$ defined in panels (a, b).}
		\label{fig:TFIM}
\end{figure} Instead of setting $n = 2p$ for producing the exact ground states~\cite{wiersema2020exploring}, our simulations focuses on configurations where $n \geq 2p + 2$—practically, $n = 2p + 2$—aiming to understand the finite-depth effects and explore the applications of finite-depth extrapolation discussed in Section~\ref{subsec:extrapolation}.

	Numerical findings are illustrated in Fig.~\ref{fig:TFIM}. Panel (a) shows the fidelity between optimized quantum states and true ground states, defined by
	\begin{equation}
		\mathcal{F}(g) = |\langle \psi(\boldsymbol{\zeta},\boldsymbol{\eta};g) | \psi_{\mathrm{gs}} \rangle|^2,
	\end{equation}
	for $p=2,3,\cdots,7$ and $n=18$ across the detection grid $\mathcal{G}_{\text{det}}$. Fidelities at the critical point $g_c = 1$ are recorded as 0.58, 0.68, 0.75, 0.80, 0.84, and 0.87, respectively. Near $g \approx 0$, fidelities fluctuate considerably due to the small spectral gap present in this region, where optimization struggles to consistently identify states close to the true ground state. Despite the optimized quantum states achieving relatively stable low energy, the fidelity to the ground state remains inconsistent. Nevertheless this region is distant from $g_c  = 1$ and minimally impacts our findings. For $g \gg 1$, even at $p=2$, the fidelities significantly increase, as the ground states are close to the initial product state. The inset of panel (a) displays the derivative of fidelity, $\partial \mathcal{F}/\partial g$, as a function of $g$, revealing that the peak does not align with the thermodynamic critical point and shifts away from it as $p$ increases. Panel (b) presents the LASSO regression's loss landscape, highlighting the minimum point which correlates with the anticipated phase transition critical point $\Tilde{g}_c$. Utilizing a relatively large regularization parameter $\lambda$ sharpens the valleys within the loss landscape, leading to the selection of a singular observable as the order parameter at the valley's lowest point. The LASSO selected order parameter here is 
	\begin{equation}
		\mathcal{O}_{\text{ml}} = \frac{1}{n}\sum_{j} \sigma^x_j\sigma^x_{j+2},
    \label{Eq: X0X2}
	\end{equation}
	which captures the long-range correlations within the system, yielding a more precise prediction of the critical point than the traditional $x$-direction magnetization, $m_x$. 
 
	The exponential decay in the error between the estimated and the exact critical point as a function of the circuit depth $p$, as predicted by our algorithm and by $m_x$, is shown in Fig.~\ref{fig:TFIM}(c,d). This suggests that, under certain assumptions (see Observation~\ref{statement:exponential_precision} in section~\ref{subsec:extrapolation}), the location of the critical point in the thermodynamic limit, $g_c$, can be predicted by applying an exponential extrapolation approach to the results for finite depth $p$, $\tilde{g}_c(p)$, i.e., $\Tilde{g}_c(p)=g_c(p\rightarrow\infty)+ c e^{-\nu p}$. By applying the equation above to the finite-$p$ critical points obtained from $m_x$, we obtain an extrapolated critical point $\Tilde{g}_c(p \rightarrow \infty) = 0.921$, with a mean square fitting error of 0.004. Moreover, using the finite-depth extrapolation technique in combination with our algorithm predicts $\Tilde{g}_c(p \rightarrow \infty) = 0.995$, yielding a mean square fitting error of 0.002, closely approximating the theoretical value of $g_c=1$. Compared with prior works \cite{dreyer2021quantum, bosse2023sketching}, our approach achieves stable and precise estimations of the critical point even with notably shallow circuits. For further evaluation, we applied polynomial fitting using the equation $\Tilde{g}_c(p)=g_c(p\rightarrow\infty)+ c p^{-\nu}$. The extrapolated $g_c(p\rightarrow\infty)$ values for $m_x$ and our algorithm are $\Tilde{g}_c(p \rightarrow \infty) = 1.566 \pm 0.003$ and $\Tilde{g}_c(p \rightarrow \infty) = 2.545 \pm 0.004$, respectively, both significantly deviating from the theoretical value of $g_c=1$, further corroborating the exponential convergence of $\Tilde{g}_c(p)$ to the thermodynamic limit value. 
	Furthermore, we compute the Pearson correlation coefficient $r$ between the estimation error $|\Tilde{g}_c - g_c|$ and the fidelity at the critical point $\mathcal{F}(g_c)$, defined as~\cite{wasserman2003statistics}:
    \begin{equation}
		r= \frac{\sum_{p}\left(|\Tilde{g}_c - g_c|_p - \overline{|\Tilde{g}_c - g_c|}\right)\left(\overline{\mathcal{F}}(g_c)-\mathcal{F}_p(g_c)\right)}{\sqrt{\sum_{p}\left((|\Tilde{g}_c - g_c|_p - \overline{|\Tilde{g}_c - g_c|}\right)^2 \sum_{p}\left(\mathcal{F}_p(g_c) - \overline{\mathcal{F}}(g_c)\right)^2}},
	\end{equation}
	\noindent $p$ indexes data from different circuit depths, and overlines denote mean values over different $p$ values. Variables exhibit strong linear correlation for $r \approx 1$, strong inverse correlation for $r \approx -1$, and no correlation for $r \approx 0$. Our analysis reveals a Pearson correlation coefficient $r=0.999$ between $1-\mathcal{F}(g_c)$ and $|\Tilde{g}_c - g_c|$ with $\Tilde{g}_c$ obtained from our loss landscape; and $r = 0.998$ between  $1-\mathcal{F}(g_c)$ and $|\Tilde{g}_c - g_c|$ with $\Tilde{g}_c$ from $m_x$.  All these results support the existence of a robust relationship:
	\begin{equation}
		|\Tilde{g}_c - g_c| \propto 1-\mathcal{F}(g_c) \propto \frac{1}{2^p}.
	\end{equation}

	\begin{figure}[tbh]
		\centering
		\includegraphics[width=8.4cm]{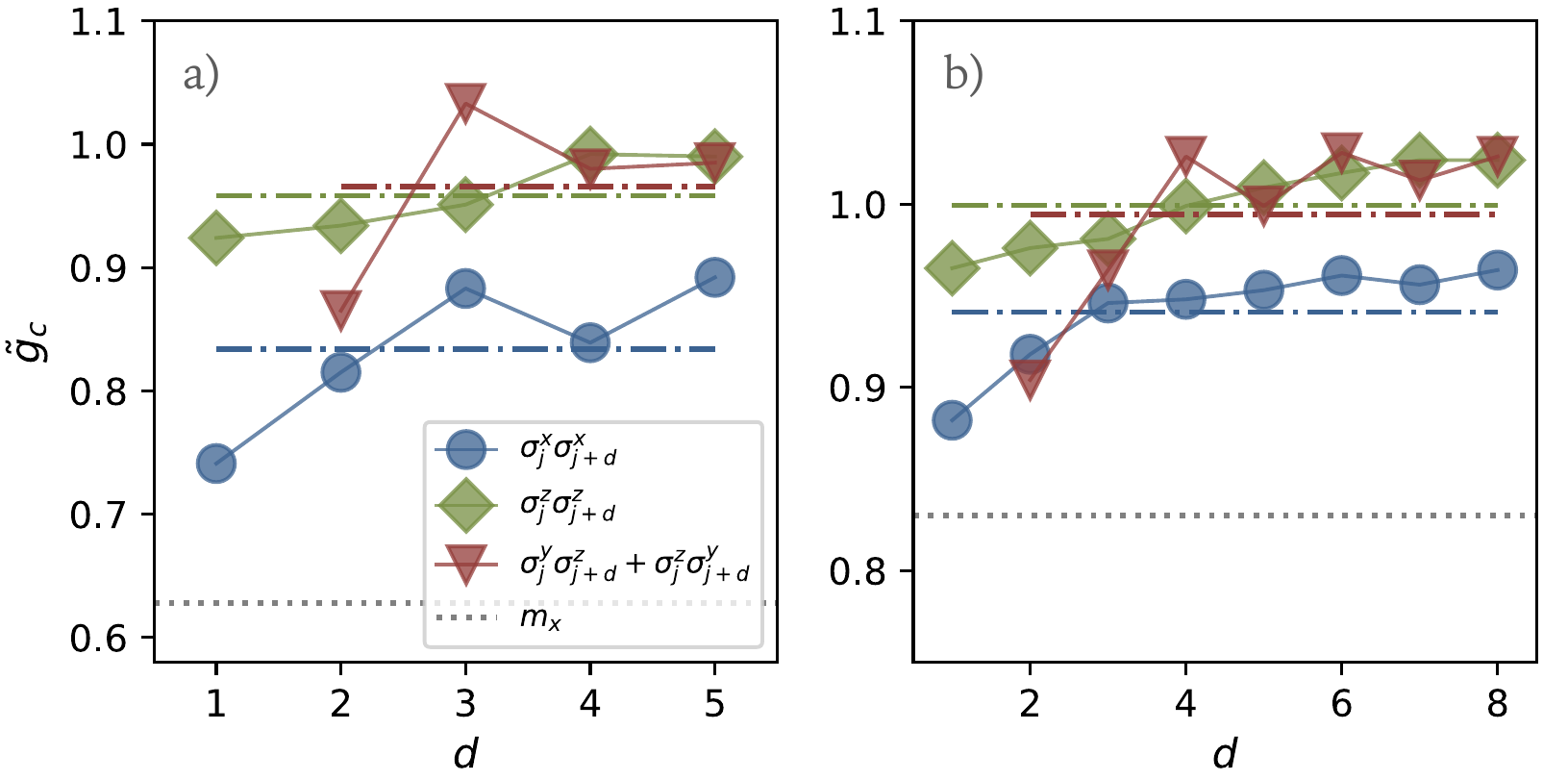}
		\caption{Estimation of critical points using various learned order parameters in the 1D transverse-field Ising model. (a) Results for optimized states at circuit depth $p=4$, with the average of $\mathcal{O}_{\mathrm{xx}}(d)$ estimating the critical point as $\Tilde{g}_c = 0.83 \pm 0.05$, the average of $\mathcal{O}_{\mathrm{zz}}(d)$ as $\Tilde{g}_c = 0.96 \pm 0.03$, and the average of $\mathcal{O}_{\mathrm{yz}}(d)$ as $\Tilde{g}_c = 0.97 \pm 0.06$. (b) Results for circuit depth $p=7$. The average of $\mathcal{O}_{\mathrm{xx}}(d)$ estimates the critical point as $\Tilde{g}_c = 0.94 \pm 0.03$, the average of $\mathcal{O}_{\mathrm{zz}}(d)$ as $\Tilde{g}_c = 1.00 \pm 0.02$, and the average of $\mathcal{O}_{\mathrm{yz}}(d)$ as $\Tilde{g}_c = 0.99 \pm 0.04$. In both panels, dashed lines indicate the average critical point estimates across different $d$ values while the dotted lines display the critical point estimate obtained from the magnetization $m_x$.}
		\label{fig:op-comparison}
	\end{figure}
    
	It is important to note that for panels (a) through (d) in Fig.~\ref{fig:TFIM}, exact calculations of measurement expectations were employed, thereby excluding potential circuit or measurement errors. To simulate the constraints encountered in experimental setups due to finite measurements, we performed 100 numerical trials. In each trial, we used a finite number of measurement shots to estimate the expectation values of local observables within $\mathcal{P}_f$. The average and the standard deviation of the predicted critical point $\Tilde{g}_c$ as a function of the number of shots is showcased in Fig.~\ref{fig:TFIM}(e). Remarkably, even with a relatively modest number of shots, approximately $5000$, the standard deviation remains minimal, as depicted by the shaded area, underscoring the protocol's robustness to shot noise and feasibility for practical quantum experiments.

    As we will discuss in section~\ref{subsec:algorithm}, our algorithm can identify the most dominant feature that serves as an order parameter to characterize the phase transition. As an interesting extension, we can iteratively refine this process by manually removing the currently identified feature, re-running the algorithm to find the next dominant feature, and repeating this process until no stable central-minima loss valley is observed. This approach allows us to identify a series of order parameters that provide estimates of the critical point. We implement this iterative approach for optimized states from quantum circuits with depths $p=4$ and $p=7$, identifying the following symmetric order parameters:
    \begin{equation}
    \begin{aligned}
        &\mathcal{O}_{\mathrm{xx}}(d) = \frac{1}{n}\sum_{j} \sigma^x_j\sigma^x_{j+d}, \\ & \mathcal{O}_{\mathrm{zz}}(d) = \frac{1}{n}\sum_{j} \sigma^z_j\sigma^z_{j+d}
    \end{aligned}
    \end{equation}
    where $d=1,2,\ldots,p+1$, and
    \begin{equation}
        \mathcal{O}_{\mathrm{yz}}(d) = \frac{1}{2n}\sum_{j} (\sigma^y_i\sigma^z_{j + d} + \sigma^z_j\sigma^y_{j+d}) 
    \end{equation}
    where $d = 2,3,\ldots,p+1$. The estimated critical points obtained from these learned order parameters over all the possible values of $d$ are shown in Fig.~\ref{fig:op-comparison}. Each order parameter provides a more precise estimate of the critical point compared to traditional $x$-magnetization. Notably, even for these states with fidelities of only 0.75 ($p=4$) and 0.87 ($p=7$) at the critical point, $\mathcal{O}_{\mathrm{zz}}$ and $\mathcal{O}_{\mathrm{yz}}$ provides an accurate estimate of the critical point when $d \geq 4$: $0.98 \lesssim \Tilde{g}_c \lesssim 1.03$ without the need for extrapolation.  Additionally, we notice that the difference between the estimated critical point $\Tilde{g}_c$ and the theoretical value $g_c = 1$ tends to diminish as $d$ increases. Furthermore, for relatively small values of $d$, extrapolation in $p$ can be employed to refine the critical point estimation.

    Additionally, we investigate the influence of noise on the optimization of the variational parameters. Our numerical simulations on the 1D TFIM (detailed in the Supplementary Material (SM)) reveal that even when the optimization is performed on noisy quantum circuits—where each 2-qubit gate is immediately followed by a local depolarizing noise channel with error rate $\epsilon^{(\text{DP})}=0.01$—the resulting optimized parameters are only marginally affected. Both the output fidelity as a function of $g$ and the LASSO loss landscapes for circuit depths $p=2$ and $p=3$ show only slight degradations compared to the noiseless case. Notably, the learned order parameter remains $\mathcal{O}_{\mathrm{ml}}=(1/n)\sum_j \sigma_j^x \sigma_{j+2}^x$, and the critical point estimates derived from the loss landscapes differ only minimally. These results indicate that our algorithm is robust to moderate hardware noise during the optimization process.
 
	\begin{figure*}[t]
		\centering
		\includegraphics[width=17cm]{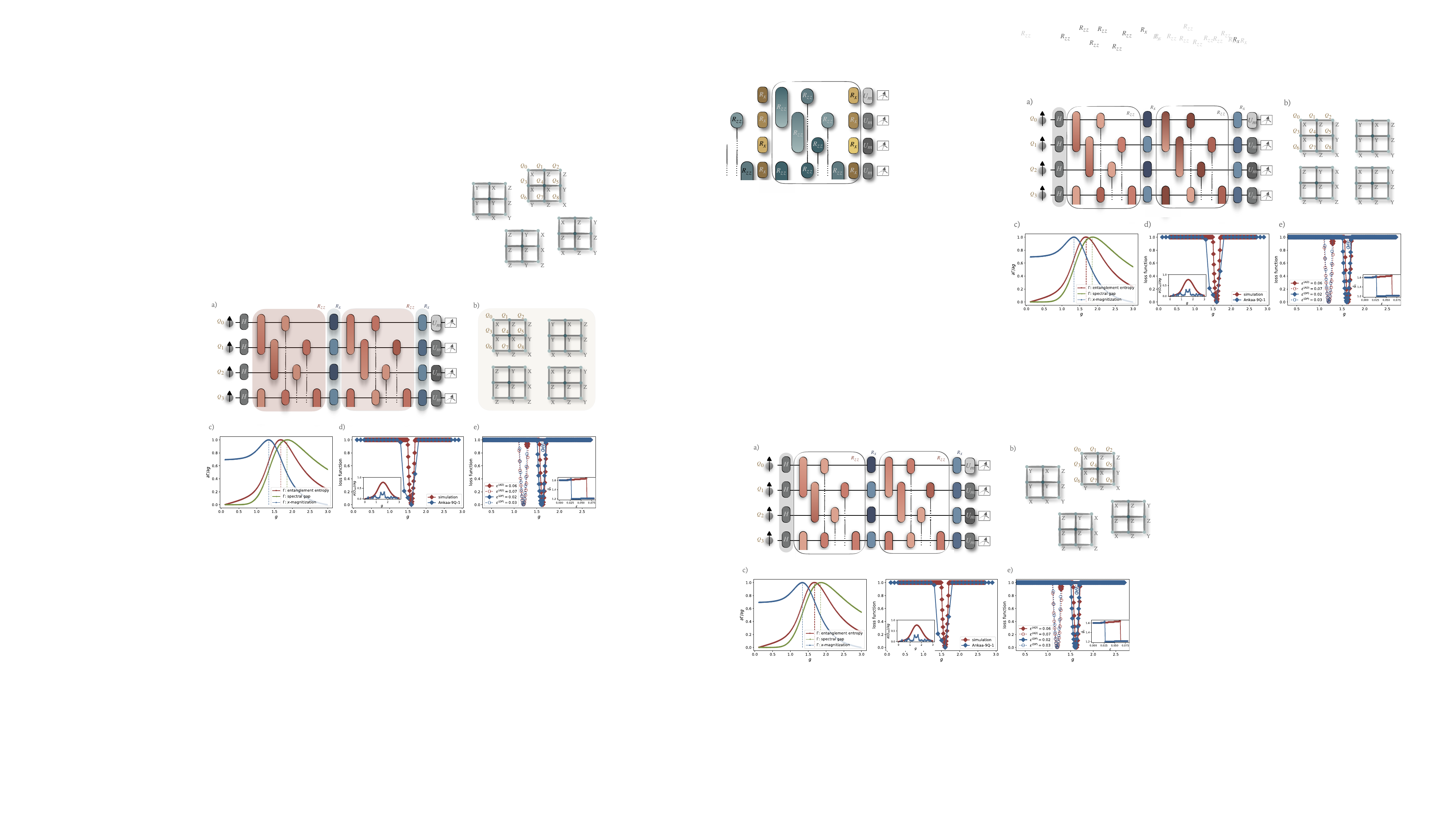}
		\caption{Analysis and results for the 2D transverse-field Ising model on a $3 \times 3$ qubit grid. (a) Quantum circuit design utilizing a tailored Hamiltonian variational ansatz that accommodates the distinct interactions within the grid. (b) Details of the four measurement protocols employed to measure the expectation values of one-weight and two-weight Pauli terms, using specific single-qubit unitaries for basis rotation before projective measurements. (c) Derivatives of the relative spectral gap and ground state entanglement entropy, hinting to a probable phase transition with a critical point between 1.68 and 1.84, contrasted against the derivative peak of $x$-magnetization at 1.34. (d) Loss minimization results from both computational and experimental analyses, pinpointing an estimate of the phase transition critical point at 1.60, which is closer to values given by relative spectral gap and entanglement entropy. The inset highlights the derivative of the learned order parameter $\langle O_{\text{ml}}\rangle$, demonstrating the algorithm's capability to predict critical points amidst experimental data variability. (e) Loss minimization results under various simulated amplitude damping (AD) and depolarizing (DP) noise rates, revealing the algorithm's robustness to noise. The inset illustrates the critical point's sensitivity to increased noise rates, with a notable shift in $\Tilde{g}_c$ values, which also is accompanied by the selection of another order parameter.}
		\label{fig:2DTFIM}
	\end{figure*}

	\subsection{2D Transverse-field Ising Model}
	\label{{Sec:2DTFIM}}
	
	Now we first analyse a 2D TFIM consisting of a $3 \times 3$ qubit grid with open boundary conditions, described by the Hamiltonian
	\begin{equation}
		\mathcal{H}_{\text{TI-2D}} = -J\sum_{\langle ij \rangle}\sigma^z_i\sigma^z_{j} - g\sum_j\sigma^x_j,
	\end{equation}
	where $\langle ij\rangle$ indicates nearest neighbour qubits. We aim to utilize the developed algorithm to accurately estimate the phase transition critical point of this system, both through numerical simulations and experimental implementation.
	
	To address the challenges posed by the varying roles of qubits and edges in the 2D lattice, we adopt a modified Hamiltonian variational ansatz. This approach specifically accounts for the distinct interactions and qubit positions as follows:
	\begin{equation}
		\begin{aligned}
			U_{\text{TI-2D}}(\boldsymbol{\beta}, \boldsymbol{\gamma})= \prod_{j=1}^p &\exp \left(-i \frac{\beta_j}{2} \mathcal{H}_{x}\right) \exp \left(-i \frac{\beta'_j}{2} \mathcal{H}'_x\right) \\ 
			&\exp \left(-i \frac{\beta''_j}{2} \mathcal{H}''_{x}\right) \exp \left(-i \frac{\gamma_j}{2} \mathcal{H}_{zz}\right) \\ 
			&\exp \left(-i \frac{\gamma'_j}{2} \mathcal{H}'_{z z}\right).
		\end{aligned}
	\end{equation}
	Here, $\mathcal{H}_{zz}$ and $\mathcal{H}'_{zz}$  correspond to the sum of $\sigma^z_j\sigma^z_{j+1}$ for edges not involving and involving the central qubit ($Q_4$), respectively. $\mathcal{H}_x$, $\mathcal{H}'_x$, $\mathcal{H}''_x$ denote the sum of $\sigma^x_j$ operators for distinct sets of qubits: $\{Q_0, Q_2,Q_6,Q_8\}$, $\{Q_1, Q_3,Q_5,Q_7\}$, and $\{Q_4\}$, respectively. This ansatz, visually depicted in Fig.~\ref{fig:2DTFIM}(a), is specifically designed to respect the inherent reflection and rotational symmetries of the lattice, ensuring that the variational parameters are symmetrically adapted to the physical layout and interaction dynamics of the system.
	
	To facilitate the estimation of expectation values for most Pauli terms in $\mathcal{P}_f$, which consist of weight one and two operators up to rotation and reflection symmetries, we implement four distinct measurement protocols, detailed in Fig.~\ref{fig:2DTFIM}(b). For qubits labeled with $X$, $Y$, and $Z$, pre-measurement single-qubit unitaries are applied to rotate the measurement basis accordingly, followed by projective measurements. This methodology ensures comprehensive analysis and measurement across the qubit array.
	
	Before implementing our algorithm across a range of $g$ values, we examine two metrics which have been used to analyse quantum phase transitions in the literature~\cite{campostrini2014finite, osborne2002entanglement, osterloh2002scaling, yuste2018entanglement}: the relative spectral gap, defined as $|(E_1 - E_{\text{gs}})/E_{\text{gs}}|$ where $E_{\text{gs}}$ represents the ground state energy and $E_1$ denotes the first excited state energy of $\mathcal{H}_{\text{TI-2D}}$, and the ground state entanglement entropy, with a focus on the entanglement between the central qubit, $Q_4$, and its complement, $Q^c_4$, within the system. This is expressed as
	\begin{equation}
		S(Q^c_4) = - \text{Tr}(\rho_{Q_4} \log_2 \rho_{Q_4}),
	\end{equation}
	where $\rho_{Q_4}$ is the reduced density matrix for $Q_4$, defined as $\rho_{Q_4} = \text{Tr}_{Q_4^c} (|\psi_{\text{gs}}\rangle \langle \psi_{\text{gs}}|)$. 
	Here, $|\psi_{\text{gs}}\rangle$  denotes the ground state of $\mathcal{H}_{\text{TI-2D}}$.
	
	In our exploration, the relative spectral gap and the entanglement entropy serve as indicators of shifts in the global properties of the ground state, as opposed to merely reflecting changes in local order parameters. Consequently, in comparison to a local order parameter such as the magnetization, these metrics are less affected by finite-size effects and emerge as more sensitive probes for detecting phase transitions within small systems. By computing the derivatives of these metrics, we identify peaks at $1.84$ and $1.68$, respectively, as illustrated in Fig.~\ref{fig:2DTFIM}(c). These peaks hint at a likely phase transition with critical point located near $1.68 \sim 1.84$.  In contrast, the derivative of the $x$-magnetization, $m_x = 1/n\sum_{j=1}^n\sigma^x_j$, exhibits a sharp decline at $1.34$, markedly deviating from the aforementioned range. Previous numerical studies have determined that the 2D TFIM undergoes a quantum phase transition at a critical point of approximately $3.04438$ in the thermodynamic limit~\cite{blote2002cluster, schmitt2022quantum}. In the infinite-size limit, predictions of the critical point from relative energy gap, entanglement entropy, and $x$-magnetization are expected to converge to this value. However, owing to current hardware constraints, our analysis in this part is confined to a $3 \times 3$ qubit system. To further validate our protocol, we extend our numerical investigations to larger system sizes with periodic boundary conditions in the SM. For the $3 \times 3$ system considered here, estimates of the critical point from the peaks in the relative energy gap and entanglement entropy are also closer to the exact value and confirm the enhanced sensitivity of these probes with respect to a local order parameter. Since obtaining a precise estimate of the critical point would require a finite-size (or depth) extrapolation, which is beyond the scope of this work, in the following we will assume the range $1.68 \sim 1.84$ as a benchmark for our protocol. If an order parameter yields an estimate that is closer to, or even surpasses, this benchmark—indicating that it converges more rapidly and is less affected by finite-size effects—then we consider that order parameter to be superior.

	Proceeding to quantum optimization, we employ a shallow quantum circuit with $p=2$. Implementing our algorithm, we carry out numerical simulations over the optimization grid points $\mathcal{G}_{\text{opt}} = \{0,0.1,0.2.,\ldots,3\}$. Then, we train the LASSO regressor over the detection grids $\mathcal{G}_{\text{det}} = \{0,0.01,0.02.,\ldots,3\}$ for numerical simulations and utilize $\mathcal{G}_{\text{det}} = \mathcal{G}_{\text{opt}}$ for experiments conducted on the Rigetti Ankaa-9Q-1 machine~\cite{rigetti2023ankaa9Q}. For the former detection grid, the fidelity between the noiseless ideal output states and the true ground states vary, ranging from $0.200$ to $0.999$, with an mean fidelity of $0.987$, indicating a minimal optimization deficit. The chosen window sizes are $w=40$ for numerical data and $w = 4$ for experimental data.

	The Ankaa-9Q-1 device's performance metrics, documented around the time of the experiment, are detailed in Table~\ref{tab:Ankaa-9Q-1_parameters}. Our initial circuit configuration includes 24 $R_{zz}$ rotation gates, 18 $R_x$ rotation gates, and 9 additional single-qubit gates preceding measurements. However, the compiled circuits exhibit variability in the quantity and type of gates, contingent upon different parameters and measurements. On average, each compiled circuit incorporates 26 CZ gates, 26 iSWAP gates, 127 $R_x$ gates, and 207 $R_z$ gates. To compute expectation values we used 30,000 shots per circuit, divided across three rounds with 10,000 shots each. It is important to highlight that in this experiment, neither gate error nor readout error mitigation techniques were applied.

	\begin{table}[ht]
		\centering
		\renewcommand{\arraystretch}{1.5}
		\begin{tabular}{l|l}
			\hline
			Parameter & Median Value \\
			\hline
			$T_1$ & \(21.18 \mu\)s  \\
			$T_2$ & \(5.3 \mu\)s  \\
			1Q RB fidelity & 99.9\% \\
			1Q sim. RB fidelity & 99.3\% \\
			2Q CZ fidelity & 97.9\% \\
			2Q iSWAP fidelity & 98.5\% \\
			Readout fidelity & \(92\%\) \\
			\hline
		\end{tabular}
		\caption{Performance metrics of the Ankaa-9Q-1 quantum processor. The single-qubit fidelities, ``1Q RB fidelity" and ``1Q sim.\ RB fidelity" are assessed through randomized benchmarking~\cite{knill2008randomized, helsen2022general} and simultaneous randomized benchmarking~\cite{gambetta2012characterization}, respectively. The latter evaluates qubit performance under conditions of cross-talk and interference that occur during the simultaneous operation of multiple qubits.}
		\label{tab:Ankaa-9Q-1_parameters}
	\end{table}

	Both numerical and experimental investigations reveal a loss minimum at 1.6, as depicted in Fig.~\ref{fig:2DTFIM}(d), offering a refinement over estimates derived from $m_x$. This also shows that our algorithm can accurately extract phase transition information without necessitating error mitigation. By integrating symmetry considerations into our assessment, we find the following learned order parameter
	\begin{equation}
		\mathcal{O}_{\text{ml}} = \frac{1}{4}\left(\sigma^x_1\sigma^x_4 + \sigma^x_3\sigma^x_4 + \sigma^x_4\sigma^x_5 + \sigma^x_4\sigma^x_7\right).
	\end{equation}
	The inset of Fig.~\ref{fig:2DTFIM}(d) showcases the derivative of $\langle \mathcal{O}_{\text{ml}} \rangle$ with respect to $g$, calculated using next-nearest values. Additionally, we assess the impact of hardware noise on $\mathcal{P}_f$, analysed under the quasi-GD noise approximation outlined in Theorem~\ref{theorem:noise_robustness}. For the purposes of our algorithm, the influence of hardware noise on $\mathcal{P}_f$ resembles a global depolarizing noise channel $\Lambda_{\epsilon}(\cdot)$ with a noise rate of $\epsilon = 0.773$. The average discrepancy between the ideal and the linearly rescaled noisy features is $0.047$.

	To further study the resilience to noise of our algorithm, we consider a scenario where each $R_{zz}$ rotation is affected by either an amplitude damping noise channel with a damping rate $\epsilon^{(\text{AD})}$ or a local depolarizing noise channel with a noise rate $\epsilon^{(\text{DP})}$. Fig.~\ref{fig:2DTFIM}(e) illustrates that the predicted phase transition point $\Tilde{g}_c$ remains approximately at $1.6$ under noise levels $\epsilon^{(\text{AD})} = 0.06$ or $\epsilon^{(\text{DP})} = 0.02$. Specifically, under amplitude damping with $\epsilon^{(\text{AD})} = 0.06$, the accumulated noise on $\mathcal{P}_f$ resembles a global depolarizing channel with a noise rate of $\epsilon = 0.317$, where the average discrepancy between the ideal and the linearly rescaled noisy features is $0.074$. For local depolarizing noise with $\epsilon^{(\text{DP})} = 0.02$, the accumulative noise on $\mathcal{P}_f$ resembles a global depolarizing channel with a noise rate of $\epsilon = 0.244$, yielding a significantly lower average discrepancy of $0.011$.  However, as the noise rate increases, $\Tilde{g}_c$ abruptly shifts to around $1.2$, indicating that higher noise levels disrupt entanglement, leading the regression model to opt for a local order parameter acting on individual, localized qubits:
	\begin{equation}
		\mathcal{O}_{\text{ml}} = \frac{1}{4}\left(\sigma^x_1 + \sigma^x_3 + \sigma^x_5 + \sigma^x_7\right),
	\end{equation}
	resulting in a significantly less accurate estimate. The inset in Fig.~\ref{fig:2DTFIM}(e) shows the relationship between $\Tilde{g}_c$ and the noise rate for both amplitude damping and depolarizing channels, with crossover points at $\epsilon^{(\text{AD})} = 0.064$ and $\epsilon^{(\text{DP})} = 0.028$, respectively. In summary, our results demonstrate that the algorithm exhibits robust performance for small to moderate noise levels—i.e., the predicted phase transition point remains nearly unchanged relative to the ideal case. The observed abrupt shift at higher noise rates is consistent with the expectation that substantial noise will fundamentally alter the quantum state, thereby affecting the performance of any algorithm.

    As discussed in the previous section, the iterative removal of identified dominant order parameters from the feature vectors facilitates the discovery of multiple additional order parameters. Table~\ref{tab:multi-op} presents the most dominant order parameter and some additional order parameters alongside the corresponding simulation and experimental results, showing a generally consistent alignment between the two.
    \begin{table}[ht]
		\centering
		\renewcommand{\arraystretch}{1.5}
		\begin{tabular}{l|l|l}
			\hline
			Additional Learned Order Parameters & Sim. & Exp.\\
            $(1/4) (\sigma^x_1\sigma^x_4 + \sigma^x_3\sigma^x_4 + \sigma^x_4\sigma^x_5 + \sigma^x_4\sigma^x_7)$ & 1.60 & 1.6  \\
            $(1/2) (\sigma^z_0\sigma^z_8 + \sigma^z_2\sigma^z_6)$ & 1.29 & 1.3  \\
            $(1/2) (\sigma^z_1\sigma^z_7 + \sigma^z_3\sigma^z_5)$ & 1.47 & 1.3  \\
            $(1/4) (\sigma^z_1\sigma^z_3 + \sigma^z_1\sigma^z_5 + \sigma^z_3\sigma^z_7 + \sigma^z_5\sigma^z_7)$ & 1.27 & 1.4  \\
			$(1/4) (\sigma^z_0\sigma^z_2 + \sigma^z_0\sigma^z_6 + \sigma^z_2\sigma^z_8 + \sigma^z_6\sigma^z_8)$ & 1.27 & 1.3  \\
            $(1/2) (\sigma^x_1\sigma^x_7 + \sigma^x_3\sigma^x_5)$ & 1.52 & 1.8  \\
			\hline
		\end{tabular}
		\caption{Comparison of predicted critical points from simulation and experimental results using different learned order parameters.}
		\label{tab:multi-op}
	\end{table}

    In the SM, we investigate the critical boundary of 2D TFIM and the performance of our protocol for larger qubit grids. Applying our algorithm to the $4 \times 4$ system yields an estimated critical point of approximately 2.45, which is significantly closer to the thermodynamic limit value.

	\subsection{Extended Su–Schrieffer–Heeger Model}
	\label{Sec:eSSH}
	Now we consider a model featuring topological phase transitions, the eSSH model~\cite{sylvain2019observation, elben2020many}. This model represents an extension of the SSH model proposed by Su, Schrieffer, and Heeger, to study topological phases within one-dimensional systems~\cite{su1979solitons}. The eSSH model incorporates additional interactions beyond the nearest-neighbour hopping and, under open boundary conditions, is characterized by the Hamiltonian
	\begin{equation}
		\begin{aligned}
			\mathcal{H}_{\mathrm{eSSH}}  = &(1-g) \sum_{j=0}^{n / 2 - 1}\left(\sigma_{2 j}^x \sigma_{2 j + 1}^x+\sigma_{2 j}^y \sigma_{2 j + 1}^y+\delta \sigma_{2 j}^z \sigma_{2 j + 1}^z\right) \\
			& +g\sum_{j=1}^{n / 2-1}\left(\sigma_{2 j - 1}^x \sigma_{2 j}^x+\sigma_{2 j - 1}^y \sigma_{2 j}^y+\delta \sigma_{2 j - 1}^z \sigma_{2 j}^z\right),
		\end{aligned}
	\end{equation}
	where $g$ modulates the interaction strength, $\delta$ denotes the anisotropy in the $z$-direction, and $n$ denotes the system size. For our studies, $n$ is set such that $n = 4q$, where $q \in \mathbb{N}^+$.

	In scenarios where the anisotropy parameter $\delta$ is small, increasing the coupling strength $g$ from 0 to 1 drives a phase transition from a trivial to a topological phase within the system~\cite{elben2020many}. Initially, the trivial phase displays a dimerized configuration, with spins forming singlet pairs predominantly with their nearest neighbours. As $g$ approaches 1, the system evolves into a topological phase, marked by the emergence of edge states that represent localized excitations at the system's boundaries. Conversely, when $\delta$ exceeds a critical threshold (approximately $\delta^\star = 1.6$), the system's phase diagram becomes more complex, maintaining the trivial and topological phases while also exhibiting a symmetry-broken phase characterized by antiferromagnetic ordering. Notably, for values of $\delta \neq 0,1$, the model is non-integrable and lacks an analytical solution; hence, numerical and heuristic methods become essential for exploring its phase behavior.

	The transitions between these phases are usually analysable through the partial reflection many-body topological invariant~\cite{pollmann2012detection}
	\begin{equation}
    \widetilde{\mathcal{Z}}_{\mathcal{R}}=\frac{\operatorname{Tr}\left(\rho_I \mathcal{R}_I\right)}{\sqrt{\left[\operatorname{Tr}\left(\rho_{I_1}^2\right)+\operatorname{Tr}\left(\rho_{I_2}^2\right)\right] / 2}},
	\end{equation}   
	where $\rho_I$ represents the density matrix of a subsystem $I=I_1 \cup I_2$, where $I_1$ includes qubits $Q_{n/4}$, $Q_{n/4+1}$, $\dots$, $Q_{n/2-1}$ and $I_2$ includes qubits $Q_{n/2}$, $Q_{n/2+1}$, $\ldots$, $Q_{3n/4-1}$,  and $\mathcal{R}_I$ is the reflection operation within $I$. Note that $\widetilde{\mathcal{Z}}_{\mathcal{R}}$ is highly non-local and non-linear. In the thermodynamic limit, $\widetilde{\mathcal{Z}}_{\mathcal{R}}$ is expected to be $-1$ in the topological phase, 0 in the symmetry-broken phase, and 1 in the trivial phase. Therefore, the  value of $\widetilde{\mathcal{Z}}_{\mathcal{R}}$ allows to identify each of the three phases.
	
	In the framework of this model, we employ the Hamiltonian variational ansatz and execute numerical optimizations through two distinct initial setups:
	\begin{itemize}
		\item Trivial phase initialization: This method begins with the initial state,
		\begin{equation}
			\left|\psi_{\mathrm{in}}\right\rangle= \bigotimes_{j=0}^{n/2-1}\left|\psi^{-}\right\rangle_{2j, 2j+1},
		\end{equation}
		where $\left|\psi^{-}\right\rangle=\frac{1}{\sqrt{2}}\left(|01\rangle-|10\rangle\right)$ and the subscripts denote qubits. This state represents the ground state of $\mathcal{H}_{\mathrm{eSSH}}$ at $g=0$ for any $\delta \geq -1$. The variational quantum circuit is then applied as follows:
		\begin{equation}
			\begin{aligned}
				U^{\text{triv}}_{\text{eSSH}}(\boldsymbol{\beta}, \boldsymbol{\gamma})= \prod_{j=1}^p &\exp \left(-i \frac{\beta_j}{2} \mathcal{H}_{xy}\right) \exp \left(-i \frac{\gamma_j}{2} \mathcal{H}_{zz}\right) \\ &\exp \left(-i \frac{\beta'_j}{2} \mathcal{H}'_{xy}\right) \exp \left(-i \frac{\gamma'_j}{2} \mathcal{H}'_{zz}\right),
			\end{aligned}
		\end{equation}
		where $\mathcal{H}_{xy}$ and $\mathcal{H}_{zz}$ are the summations of $\sigma^x_{j} \sigma^x_{j+1} + \sigma^y_{j} \sigma^y_{j+1}$ and $\sigma^z_{j} \sigma^z_{j+1}$ on odd edges, respectively, and $\mathcal{H}'_{xy}$ and $\mathcal{H}'_{zz}$ are their counterparts on even edges.
		\item Topological phase initialization: This approach starts with 
		\begin{equation}
			\left|\psi_{\mathrm{in}}\right\rangle=\left|\psi^{-}\right\rangle_{0, n-1} \bigotimes_{j=0}^{n/2-2}\left|\psi^{-}\right\rangle_{2j+1, 2j+2},
		\end{equation}
		corresponding to one of the four ground states of $\mathcal{H}_{\mathrm{eSSH}}$ at $g=1$ for any $\delta \geq -1$. The variational quantum circuit for this setup is applied as:
		\begin{equation}
			\begin{aligned}
				U^{\text{topo}}_{\text{eSSH}}(\boldsymbol{\beta}, \boldsymbol{\gamma})= \prod_{j=1}^p &\exp \left(-i \frac{\beta'_j}{2} \mathcal{H}'_{xy}\right) \exp \left(-i \frac{\gamma'_j}{2} \mathcal{H}'_{zz}\right) \\ &\exp \left(-i \frac{\beta_j}{2} \mathcal{H}_{xy}\right) \exp \left(-i \frac{\gamma_j}{2} \mathcal{H}_{zz}\right),
			\end{aligned}
		\end{equation}
		where $\mathcal{H}_{xy}$, $\mathcal{H}_{zz}$, $\mathcal{H}'_{xy}$, and $\mathcal{H}'_{zz}$ are defined as above.
	\end{itemize}

\begin{figure*}[tbh]
		\centering
		\includegraphics[width=17cm]{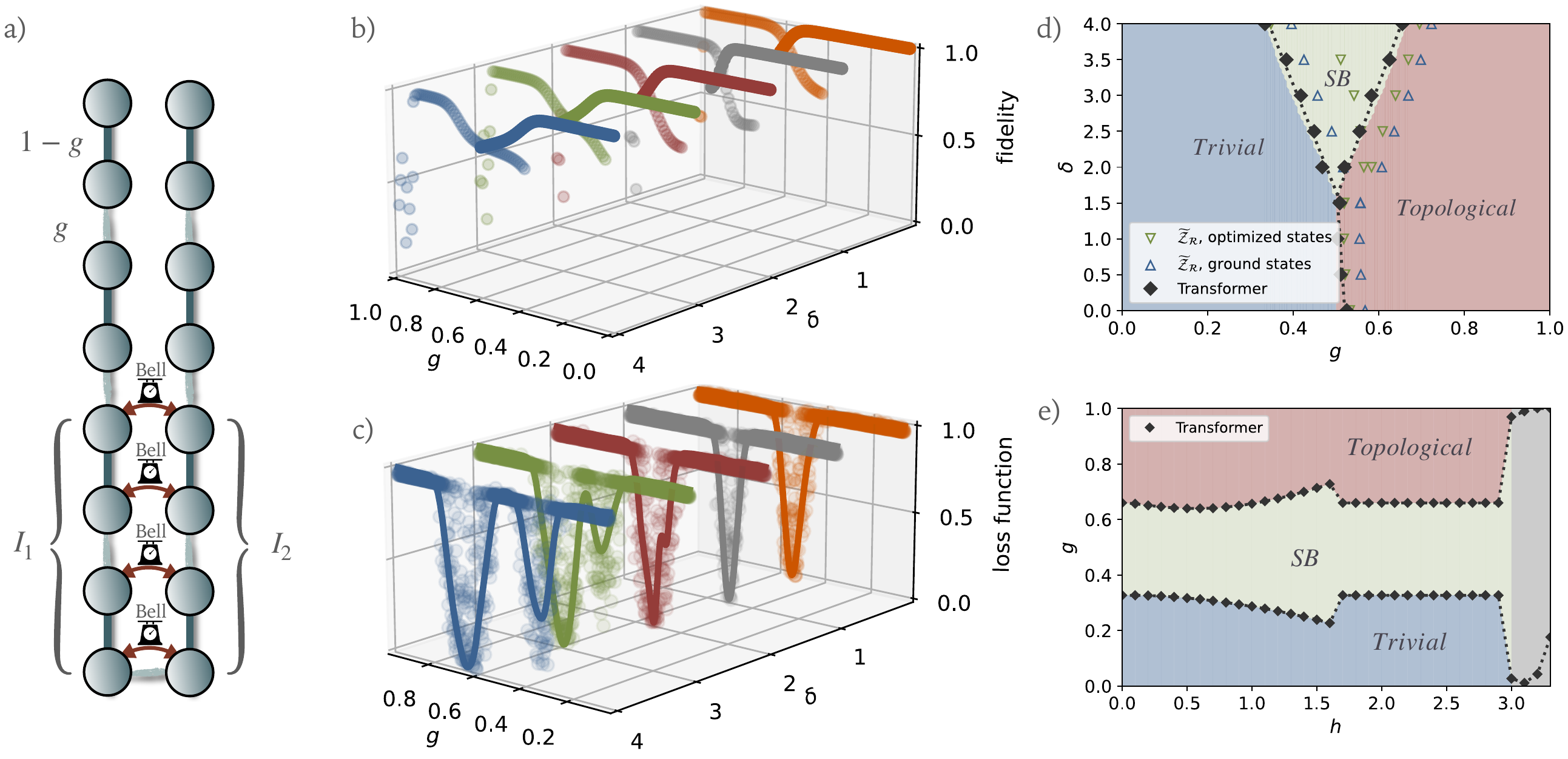}
		\caption{Overview of phase transition detection for the extended Su-Schrieffer-Heeger (eSSH) model. (a) Illustration of the eSSH model's interaction scheme and the procedure for conducting joint Bell measurements on symmetrically positioned qubit pairs across subsystems $I_1$ and $I_2$ to derive the feature vector.  (b) Fidelity comparison between the optimized quantum states and the true ground states, evaluated across various $g$ and $\delta$ values. (c) Plot of both the original and Gaussian-filter-smoothed loss landscapes for various $\delta$ values, underscoring the identification of critical points. Note that for $\delta = 2,3,4$ two minima can be identified, signalling the presence of the two possible phase transitions between the trivial and topological phases, and between the trivial and symmetry-broken ones. (d) Comparison between the predicted phase boundaries, represented by green and blue hollow triangles (indicative of transitions computed from optimized and ground states using the partial-reflection many-body topological invariant $\widetilde{\mathcal{Z}}_{\mathcal{R}}$), and solid black diamonds representing the phase boundaries predicted by the Transformer model. This comparison is set against the phase diagram from Ref.~\cite{meth2022probing}, derived numerically using the infinite-size density matrix renormalization group technique. (e) Application of learned order parameters to the modified eSSH model with a transverse field ($\mathcal{H}_{\text{T-eSSH}}$), where $\delta = 4$. This panel tracks the phase transitions as the field strength $h$ varies, with phase transitions identified where the Transformer model predicts shifts in phase labels, showcasing the adaptability of the learned parameters across different physical scenarios.}
		\label{fig:eSSH}
	\end{figure*}

	The optimization grid for the trivial phase initialization spans from $g_{\text{min}} = 0$ to $g_{\text{max}} = 0.6$, with a sampling resolution of $\delta g = 0.05$. Conversely, the grid for the topological phase initialization ranges from $g_{\text{min}} = 0.4$ to $g_{\text{max}} = 1$, with the same sampling resolution of $\delta g = 0.05$. Given the emphasis on symmetry-oriented phase transitions within this investigation, we first consider a scenario where $\mathcal{P}_f$ includes all non-identity, reflection-symmetric Pauli operators on subsystem $I$, detailed as follows:
	\begin{equation}
		\mathcal{P}_f=\left\{\begin{array}{l}
			\sigma_{n / 2 - 1}^x \sigma_{n / 2}^x, \\
			\cdots \\
			\sigma_{n / 2-2}^y \sigma_{n / 2 - 1}^z \sigma_{n / 2}^z \sigma_{n / 2+1}^y \\
			\cdots \\
			\sigma_{n / 4}^x \sigma_{n / 4+1}^z \sigma_{n / 4+2}^y \ldots \sigma_{3 n / 4-3}^y \sigma_{3 n / 4-2}^z \sigma_{3 n / 4-1}^x, \\
			\ldots
		\end{array}\right\}.
	\end{equation}
	The simultaneous acquisition of their expectation values is facilitated through the execution of joint Bell measurements on pairs of qubits within $I$ that are symmetrically positioned, such as $(Q_{n/4}, Q_{3n/4-1})$, $(Q_{n/4+1}, Q_{3n/4-2})$, and so forth, up to $(Q_{n/2-1},Q_{n/2})$. Fig.~\ref{fig:eSSH}(a) displays these joint Bell measurements performed on the corresponding qubit pairs in subsystems $I_1$ and $I_2$, illustrating the method for deriving the feature vector for a 16-qubit quantum chain. Notably, in the absence of noise, classically computing expectation values of such large-weight Pauli operators in $\mathcal{P}_f$ becomes significantly more resource-intensive, indicating a possible avenue towards practical quantum advantage.

	To implement our algorithm, we employ a Set Transformer-based regressor, described in Section~\ref{subsec:algorithm}. This model is configured with four attention heads and operates at a learning rate of 0.001. Initially, the model applies a self-attention mechanism to the input features to discern dependencies and relationships. Following this, the architecture incorporates a predictive layer that performs a linear transformation, mapping the input features to an intermediate vector of dimension 128. This transformation is augmented by the ReLU activation function, which introduces non-linearity, and is followed by another linear transformation that reduces the feature dimensionality to produce a single output label. Here, the detection grid extends from $\hat{g}_{\text{min}} = 0$ to $\hat{g}_{\text{max}} = 1$, with a sampling resolution of $\delta \hat{g} = 0.001$. The analysis employs a window size of $w = 50$. The overall loss function landscape is constructed by integrating the losses from the trivial phase initialization for $g \leq 0.5$ with those from the topological phase initialization for $g > 0.5$, for different values of $\delta$. Considering the notable volatility observed in the loss landscape, a Gaussian filter is applied to smooth the curve effectively. It is important to note that while the mathematical form of the order parameter learned here is not straightforwardly intuitive, it is encoded within the weights of the Transformer's neural network layers. This order parameter can be readily calculated from measurement results using the recorded parameters of the Transformer, facilitating practical applications and analyses. 
	
	For the eSSH model, we sample a single set of initial parameters for optimization, fixing $n = 16$, and circuit depth $p=5$. The measurement scheme and numerical results of our method are presented in Fig.~\ref{fig:eSSH}. Specifically, panel (b) shows the fidelities between the optimized states and the true ground states across a range of $g$ values for $\delta = 0,1,2,3,4$. Here, solid circles represent states from the trivial phase initialization (with $g\in[0, 0.6]$), while hollow circles indicate states from the topological phase initialization (with $g\in[0.4, 1]$). Notably, for states from the trivial phase initialization, the fidelity achieves its maximum near $g=0$, then decreases as $g$ increases. Conversely, states from the topological phase initialization exhibit high fidelity at relatively large $g$ values, although fidelity near $g=1$ can still be small due to the diminishing spectral gap. Panel (c) displays the original and smoothed loss landscapes for $\delta = 0,1,2,3,4$. 
	Notably, for $\delta = 0,1$, there is a single trivial-topological phase transition, while for $\delta = 2,3,4$, two phase transitions are observed, including a symmetry-broken phase, in accordance with the known phase diagram of the eSSH model~\cite{elben2020many}.

	The shaded areas in panel (d) of Fig.~\ref{fig:eSSH} reflect the phase diagram as reported in Ref.~\cite{meth2022probing}, which was calculated using the partial-reflection many-body topological invariant $\widetilde{\mathcal{Z}}_{\mathcal{R}}$ employing the infinite-size density matrix renormalization group technique~\cite{schollwock2011the}. We compute $\widetilde{\mathcal{Z}}_{\mathcal{R}}$ for both the optimized and true ground states of $\mathcal{H}_{\mathrm{eSSH}}$ at $n=16$. For $\delta < 1.6$, the transition between the trivial and topological phases is marked where $\widetilde{\mathcal{Z}}_{\mathcal{R}} \approx 0$; for $\delta > 1.6$, the transition between the trivial and symmetry-broken phases is identified where $\widetilde{\mathcal{Z}}_{\mathcal{R}} \approx 0.5$, and between the symmetry-broken and topological phases where $\widetilde{\mathcal{Z}}_{\mathcal{R}} \approx -0.5$. Green and blue hollow triangles represent the predicted phase boundaries from optimized and ground states, respectively, showing significant variance from those calculated using the partial-reflection invariant due to finite-size effects. Grey solid diamonds indicate the phase boundaries predicted by the Set Transformer for $\delta = 0, 0.5, 1, 1.5, \dots, 4$, demonstrating a good agreement with the calculated ones. The small discrepancy is mainly due to residual finite-size effects. It is important to note that our method does not presuppose the number or location of phase transitions. Instead, the presence (or absence) of valleys in the loss landscape—observed during the variational optimization—serves as an indicator of phase transitions. Therefore, the tailored ansätze are employed solely to facilitate efficient convergence in the respective regimes, without relying on any prior knowledge of the eSSH model’s correct phase structure.
	
	These results suggest that our methodology effectively learns an order parameter with minimal finite-size effects. Once learned, these order parameters (encoded within the Transformer's weights where the loss function is minimal) can be further applied to various other quantum systems or models, particularly those within the same family or those exhibiting similar phase transition characteristics. As an example, we now focus on the eSSH model with $\delta = 4$ and in the presence of a transverse field with strength $h$,
	\begin{equation}
		\mathcal{H}_{\text{T-eSSH}} = \mathcal{H}_{\mathrm{eSSH}}(\delta, g) - h\sum_j \sigma^x_j.
	\end{equation}
	We reuse the two order parameters learned in the previous case with $h=0$ to locate transitions between the trivial and symmetry-broken phases, and between the symmetry-broken and topological phases within this extended model.

Using exact diagonalization, we generate the ground states and corresponding feature vectors of $\mathcal{H}_{\text{T-eSSH}}$ for various $h$ and $g$ values, fixing $\delta = 4$. We apply the recorded Transformer order parameters to learn the phase diagram, depicted in panel (e) of Fig.~\ref{fig:eSSH}. Phase transitions are estimated at points where the Transformer model predicts a label of 0, indicative of shifts between phases. Notably, when $h < 1.7$, the phase boundaries vary smoothly as a function of $h$; then, a sudden change occurs at $h=1.7$, and between $1.7 \leq h < 3$, the boundaries are relatively stable with respect to $g$. For $h \geq 3$, no stable phase transitions can be detected, indicating that the phase diagram is dominated by a single phase with respect to our order parameter in this regime.

\begin{figure}[tbh]
		\centering
		\includegraphics[width=8cm]{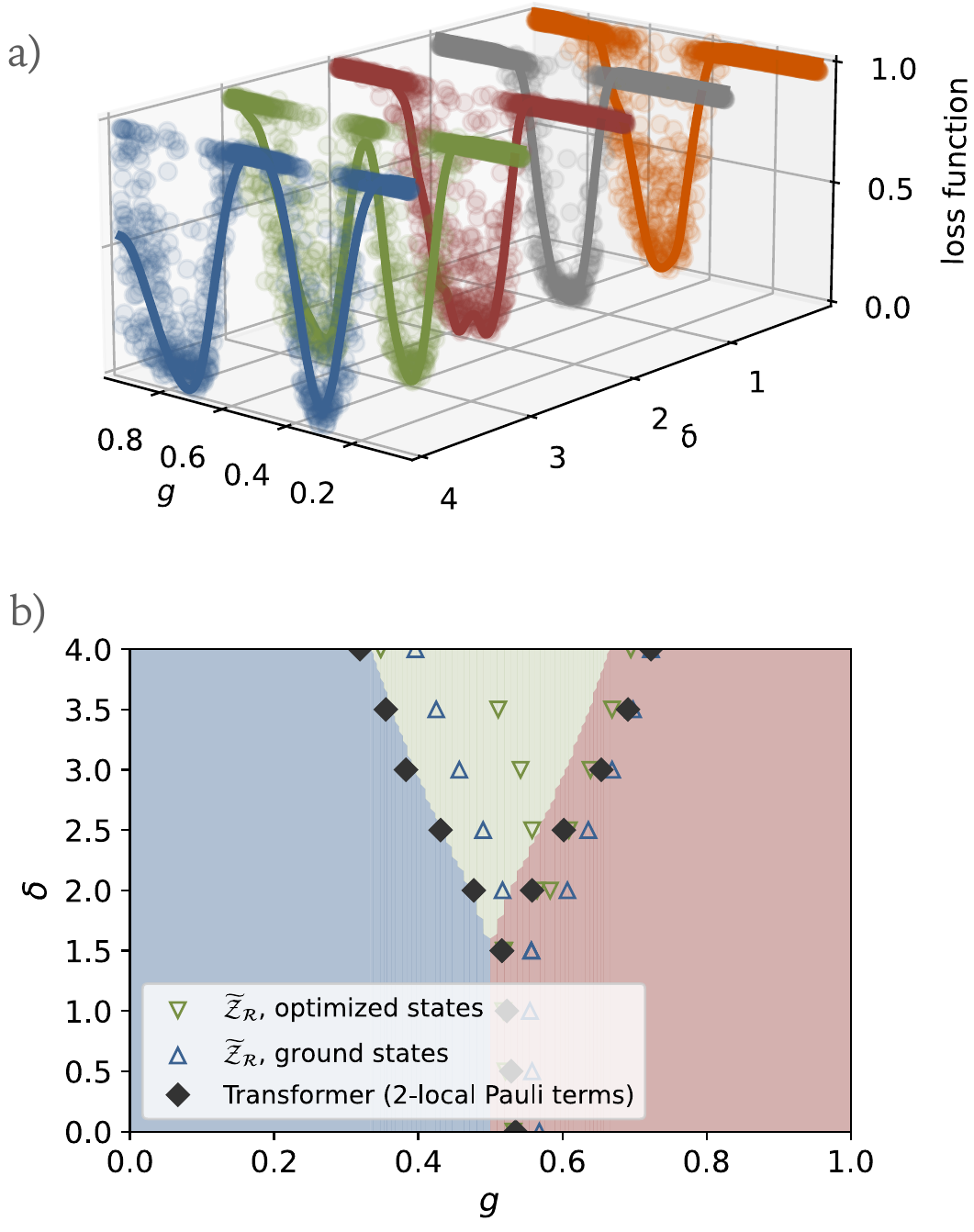}
		\caption{(a) Plot of the original and Gaussian-filter-smoothed loss landscapes constructed by applying the Transformer model using expectation values of all local Pauli operators with weight below three for various $\delta$ values. For $\delta = 2, 3, 4$, two minima are identified. (b) Comparison between the predicted phase boundaries, represented by solid black diamonds (Transformer model predictions from $2$-local Pauli terms), green hollow triangles (optimized states), and blue hollow triangles (ground states), against the phase diagram from Ref.~\cite{meth2022probing}, which was derived numerically using the infinite-size density matrix renormalization group technique.}
		\label{fig:eSSH2}
	\end{figure}

Next, we consider an alternative scenario in which no prior knowledge of the eSSH model is assumed. In this case, we construct $\mathcal{P}_f$ by including all local Pauli operators with weight two or less.   Employing the same hyperparameters as described earlier, our methodology retains its ability to detect the critical phase boundaries and identify the emergence of the symmetry-broken phase. The Transformer loss landscapes are depicted in Fig.~\ref{fig:eSSH2}(a). They reveal two distinct minima for $\delta = 2, 3, 4$, corresponding to the transitions between the trivial and symmetry-broken phases, and between the symmetry-broken and topological phases. The predictions remain largely consistent with the known phase boundaries, although finite-size effects become more pronounced in this setup, and the overall accuracy experiences a slight reduction, as shown in Fig.~\ref{fig:eSSH2}(b).  These results suggest that the non-linear functions of 2-local Pauli terms learned by the Transformer model are sufficient to characterize the topological phase transitions in the eSSH model, even without explicit prior knowledge of the system's properties.

\section{Discussion}\label{sec:conclusions}
In this work, we have presented a hybrid quantum optimization-machine learning algorithm designed to identify phase transitions in quantum systems, leveraging locally trapped states produced by shallow variational circuits instead of exact ground-state data. This approach is particularly significant in the context of near-term quantum devices, where noise and circuit depth limitations preclude the accurate preparation of exact ground states. Through numerical simulations and real-hardware experiments of the 1D and 2D transverse-field Ising models (TFIMs) and the extended Su-Schrieffer-Heeger (eSSH) model, we have demonstrated the algorithm's robustness and precision under the tested conditions.

Our real-hardware experiments validate the performance of our method under realistic noise conditions and represent, to our knowledge, one of the first experimental identifications of such order parameters directly from quantum data. Our findings illustrate that the algorithm not only detects the critical points of phase transitions but also unveils novel order parameters with faster convergence rates toward the thermodynamic limit and reduced sensitivity to finite-size effects in the systems investigated. Specifically, the LASSO algorithm identifies physically meaningful order parameters that enhance interpretability, while the Transformer model captures complex, non-intuitive many-body topological invariants that—although less directly interpretable—provide useful insights into phase transitions. Moreover, we have developed a new global optimization strategy that generates a continuous set of quantum states as a function of the Hamiltonian parameter $g$, thereby allowing precise tracking of phase evolution without requiring significantly deeper circuits.

We acknowledge that the system sizes investigated in this study are relatively modest, due to current hardware limitations and the need to measure numerous observables for training. While the results are encouraging, we are not claiming a formal quantum advantage here. Instead, the capacity of near-term devices to measure multiple observables in parallel suggests a potential practical speedup, especially if deeper or more adaptive measurements are employed. It is worth noting that while previous studies such as Herrmann \textit{et al.}~\cite{Herrmann2022Realizing} and Cho \textit{et al.}~\cite{Cho2024Machine} have employed machine learning on experimental data, our approach addresses the more demanding task of identifying unknown quantum phase transitions using shallow-circuit variational optimization under realistic noise conditions.

An open question is the extent to which our approach—particularly in the shallow-circuit or local-observable regime—remains classically efficiently simulable~\cite{Cerezo2024Does, Bermejo2024Quantum}. While shallow circuits with limited entanglement can, in some cases, be simulated using classical methods such as tensor networks~\cite{Zhou2020What, Cheng2021Simulating, Wahl2023Simulating}, our algorithm is not inherently restricted to such settings. A central finding of our study is that precision improves exponentially with circuit depth, enabling accurate results with circuits of depth $O(\log n)$; notably, such circuits are not necessarily classically simulable. Finally, because our protocol does not require the ansatz to reach the exact ground state at every value of Hamiltonian parameter, we can use a curriculum approach: optimize the circuit in an easy, barren plateau‑free region~\cite{Cao2024Exploiting}, then transfer and fine‑tune the parameters as we move into classically intractable regimes.

Furthermore, being ‘classically simulable in principle’ rarely translates into practical feasibility. In many scenarios, the required algorithms suffer from large polynomial overheads or complex data structures that become prohibitive as $n$ grows~\cite{Cerezo2024Does}. Even if a shallow-circuit model is formally simulable, its resource demands may still outstrip available classical computing power for realistic system sizes. Mid-circuit measurements~\cite{Fossfeig2023Experimental, Watts2025Quantum} can further complicate classical strategies, and integrating our algorithm with established quantum-advantage protocols or fault-tolerant schemes~\cite{Hangleiter2023Computational, Gyurik2024Exponential} may push shallow-depth training into regimes known to be classically intractable. While some configurations may appear classically tractable, the precise boundary between classical and quantum regimes remains unsettled, even in the shallow-depth, low-weight observable context explored here. Our results offer new insights into this ongoing debate.

Future improvements to our algorithm could explicitly focus on analysing differences between feature sets, rather than individual features alone. This strategy could enhance our understanding of transition dynamics, especially in systems where order parameters are not well-defined or universally applicable, such as in first-order phase transitions. Moreover, investigating our algorithm's performance within the finite-temperature critical region surrounding quantum critical points represents a promising research direction. This exploration could provide deeper insights into how quantum fluctuations, thermal effects, and optimization deficits interact at criticality, thereby enriching our understanding of finite-temperature phase diagrams.

Our study leverages classical machine learning techniques to detect quantum phase transitions from quantum data. The efficiency of the quantum variational optimization subroutine, crucial for data acquisition, can be significantly enhanced by employing strategies such as parallelism and joint Bell measurements~\cite{mineh2023accelerating, cao2024accelerated}. Given recent successes of quantum machine learning in phase recognition~\cite{wu2023quantum, Wang2024Quantum}, it is compelling to further explore quantum-enhanced methods specifically aimed at detecting quantum phases. Importantly, while we do not claim an unconditional quantum advantage for shallow-circuit regimes, the capacity to gather complex or global observables (e.g., multi-qubit correlators) will ultimately surpass classical capabilities in practice—particularly as hardware improves and deeper circuits become viable. Questions regarding the limitations of classical machine learning in this context, and whether quantum machine learning could offer an advantage, are ripe for investigation. The further exploration of these questions could open new applications of machine learning in quantum physics.

\section{Methods}\label{sec:general_framework}
	In this section, we outline the classical machine learning subroutines implemented in our study, describe the algorithm developed for detecting and locating phase transitions, and detail the extrapolation protocols designed to accurately determine the critical point of phase transition in the thermodynamic limit.
	
	\subsection{Machine Learning Preliminaries}\label{subsec:machine_learning}
	
	\paragraph{LASSO for order parameter selection.} The Least Absolute Shrinkage and Selection Operator (LASSO) algorithm is a regression technique in machine learning that incorporates both variable selection and regularization to enhance the prediction accuracy and interpretability of statistical models~\cite{tibshirani1996regression}. The target cost function for our LASSO application is formulated as:
	\begin{equation}\label{eq:Lasso_loss}
		\mathcal{C}(\boldsymbol{\kappa}, \lambda) = \left(\frac{1}{4 w} \sum_{i=1}^{2 w}\left(l_i-\kappa_0-\sum_{j=1}^{\ell} \kappa_j f_{i j}\right)^2+\lambda \sum_{j=1}^{\ell}\left|\kappa_j\right|\right),
	\end{equation}
	where $2w$ is the number of feature vectors and $\ell$ represents the number of features within each vector. The label $l_i$ (either $-1$ or $+1$) categorizes the phase associated with the $i$-th feature vector. Each component $f_{ij}$ of the feature vector corresponds to the expectation value of the $j$-th Pauli operator ($O_j$) from a predefined Pauli set,
	\begin{equation}
		\mathcal{P}_f = \{O_1, O_2, \ldots, O_\ell\},
	\end{equation}
	and calculated from the optimized quantum state $\rho_i$. The feature vector for each state is thus:
	\begin{equation}
		\boldsymbol{f}_i = \left(\operatorname{Tr}(O_1 \rho_i), \operatorname{Tr}(O_2 \rho_i), \ldots, \operatorname{Tr}(O_\ell \rho_i) \right),
	\end{equation}
	which captures essential characteristics critical to distinguishing phase properties. The coefficients $\kappa_j$ are dynamically adjusted during the optimization, where $\lambda$ is a non-negative regularization parameter that imposes a penalty on the magnitude of the coefficients. This penalty encourages sparsity in the model, enhancing interpretability by emphasizing only the most significant features, thus simplifying the model and aiding in the identification of key order parameters.

	In our implementation, LASSO is employed to discover simpler and physically meaningful order parameters, enabling direct identification of critical physical quantities. By adaptively adjusting the $\lambda$ parameter, our approach finely tunes the width of the valleys within the classical loss landscape—the larger the value of $\lambda$, the narrower the valleys—thereby avoiding issues of overly broad or narrow valleys, and ensuring the retention of essential features.

	\vspace{1em}
	
	\paragraph{Transformers for complex order parameter synthesis.} 
	The Transformer neural network, developed by Vaswani \textit{et al.} in 2017, significantly advances the handling of sequential data with its principal component: the self-attention mechanism, also known as scaled dot-product attention~\cite{vaswani2017attention}. This mechanism enables the Transformer to dynamically prioritize different segments of the input feature vector \(\boldsymbol{f}_i\) based on the relevance of different features, quantified as
	\begin{equation}
		\operatorname{Attention}(\mathbf{Q}_i, \mathbf{K}_i, \mathbf{V}_i) = \operatorname{softmax}\left(\frac{\mathbf{Q}_i \mathbf{K}_i^T}{\sqrt{d_k}}\right) \mathbf{V}_i,
	\end{equation}
	where $\mathbf{Q}_i$, $\mathbf{K}_i$, and $\mathbf{V}_i$ represent the queries, keys, and values respectively, each derived from the input feature vectors \(\boldsymbol{f}_i\). Here, $d_k$ is the dimension of the keys. Queries highlight the current focus within the input, keys facilitate the alignment of these queries with relevant data points, and values convey the substantial data intended for output. The  attention formula enables the Transformer to adjust attention across the features dynamically, which enhances its ability to discern complex dependencies. This capability may be particularly valuable in detecting quantum phase transitions, as it not only identifies critical observables but also elucidates the intricate dependencies among their expectation values, thereby enabling a comprehensive understanding and synthesis of observables related to the phase transitions.

	The multi-head attention concept~\cite{vaswani2017attention} further extends this capability, enabling the model to dynamically assign different weights of significance to various segments of the input data. This is achieved through the multi-head attention,
	\begin{equation}
		\operatorname{MultiHead}(Q, K, V) = \operatorname{Concat}(\operatorname{head}_1, \ldots, \operatorname{head}_h) W^O,
	\end{equation}
	where each head, $\operatorname{head}_j$, performs attention operations independently:
	\begin{equation}
		\operatorname{head}_j = \operatorname{Attention}(Q W_j^Q, K W_j^K, V W_j^V),
	\end{equation}
	using distinct weight matrices $W_j^{Q}, W_j^{K}, W_j^{V}$, and $W^O$. The $W^O$ matrix is a final linear transformation applied to the concatenated outputs of all heads before producing the final output. This design enables the Transformer to capture a richer representation by focusing on different aspects of the input feature vectors in parallel.

	\begin{figure}[b]
		\centering
		\includegraphics[width=7cm]{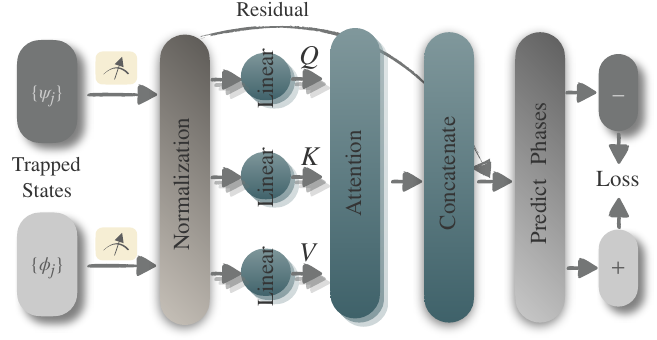}
		\caption{Schematic depiction of the Set Transformer-based regression model tailored for detecting phase transitions. The diagram methodically presents the model's architecture, incorporating key processes like input state measurement processing, feature normalization, the generation of $Q$, $K$, and $V$ through linear weight matrices, scaled dot-product attention, concatenation of multiple heads, a residual connection, a prediction layer for determining phase labels.}
		\label{fig:schematic_st}
	\end{figure}

	Given the non-sequential nature of the input features in our study—labeled either $1$ or $-1$ without a dependence on sequence—the Set Transformer, adapted by Lee \textit{et al.}~\cite{chaudhuri2019set}, offers an ideal framework for our application. The Set Transformer excels at capturing complex interrelations within data sets by employing a series of computational stages:
	\begin{enumerate}
		\item \textbf{Normalization}: The input feature vectors $\{\boldsymbol{f}_i\}$ undergo layer normalization to reduce internal covariate shift, enhancing learning stability.
		\item \textbf{Multi-head attention}: The normalized feature vectors $\{\Tilde{\boldsymbol{f}}_i\}$ are processed with multi-head attention, allowing the model to concurrently analyse multiple features of the input set. Output from this stage includes a dropout step to prevent overfitting.
		\item \textbf{Residual connection}: Integrates the original feature vector with the attention output to preserve gradient flow, enhancing training effectiveness.
		\item \textbf{Prediction layer}: Finally, the attention-enhanced data passes through a fully connected layer with ReLU activation, where the final phase labels are predicted.
	\end{enumerate}
	
	We utilize the Adam optimizer~\cite{kingma2017adam} with an appropriate learning rate to optimize parameters within this framework, aiming to minimize the mean-square loss between the predicted phase labels and the assigned ones. Through its comprehensive handling of feature interdependencies, the Set Transformer exhibits improved performance in our benchmarks with complex and non-local order parameters. The architectural details and operational functionalities of our Set Transformer-based regression model are schematically depicted in Fig.~\ref{fig:schematic_st}. An exemplary implementation of the Set Transformer architecture used in our experiments, along with its corresponding complexity analysis, is provided in the SM.

	\vspace{1em}
	In this work, we employ a LASSO-based algorithm to detect phase transitions between the ferromagnetic and paramagnetic phases within 1D and 2D TFIMs in Sections \ref{Sec:1DTFIM} and \ref{{Sec:2DTFIM}}, respectively, and utilize a Transformer-based algorithm for identifying topological phase transitions among trivial, symmetry-broken, and topological phases in the eSSH model in Section \ref{Sec:eSSH}.

	\subsection{Algorithm}\label{subsec:algorithm}
	
	\paragraph{Quantum local traps} The first stage of our protocol is the acquisition of locally optimized quantum states through variational optimization~\cite{cerezo2021variational}. We focus on a specific parameterized Hamiltonian $\mathcal{H}(g)$, where $g$ is selected from the interval $[\hat{g}_{\text{min}}, \hat{g}_{\text{max}}]$. 
	
	To systematically explore this range, we establish an optimization grid, $\mathcal{G}_{\text{opt}}$, defined by points $\{g_{\text{min}}, g_{\text{min}} + \delta g, g_{\text{min}} + 2\delta g,\ldots, g_{\text{max}}\}$, where $\delta g$ represents the optimization sampling resolution. Here, $g_{\text{min}}$ is typically set to be smaller than $\hat{g}_{\text{min}}$ but it might also be slightly greater, while $g_{\text{max}}$ is generally larger than $\hat{g}_{\text{min}}$ but it could also be marginally smaller. Introducing points outside the primary interval $[\hat{g}_{\text{min}}, \hat{g}_{\text{max}}]$ in $\mathcal{G}_{\text{opt}}$ helps to mitigate the risk of overfitting during the optimization process. When the ground state manifold exhibits relatively low complexity, it is feasible to employ a relatively large sampling resolution. 
	
	Subsequently, we introduce a variational quantum circuit $U(\boldsymbol{\gamma}, \boldsymbol{\beta};g)$ and apply the Fourier series method for its optimization, a strategy described by Zhou \textit{et al.}~\cite{zhou2020quantum}. For a $p$-layer quantum circuit, we express certain rotation angles within the $j$-th layer as 
	\begin{equation}
		\gamma_j=\sum_{k=1}^p \Tilde{\gamma}_k \sin \left(\Big(k-\frac{1}{2}\Big)\Big(j-\frac{1}{2}\Big) \frac{\pi}{p}\right),
	\end{equation}
	with the sine function being used. Conversely, the remaining rotation angles are written as
	\begin{equation}
		\beta_j=\sum_{k=1}^p \Tilde{\beta}_k \cos \left(\Big(k-\frac{1}{2}\Big)\Big(j-\frac{1}{2}\Big) \frac{\pi}{p}\right),
	\end{equation}
	with the cosine function being employed. Considering the Hamiltonian variational ansatz for the transverse-field Ising model, as explored in Refs.~\cite{wecker2015progress, wiersema2020exploring}, we designate the $R_{zz}$ rotation angle in the $j$-th layer as $\gamma_j$, and the $R_{x}$ rotation angle in the same layer as $\beta_j$. This approach to angle assignment can naturally generalize to variational ansätze that incorporate more parameters within each layer. The variational Fourier coefficients, ${\Tilde{\gamma}_k}$ and ${\Tilde{\beta}_k}$, play a pivotal role in this framework. In the update of ${\Tilde{\gamma}_k}$ and ${\Tilde{\beta}_k}$, each coefficient is written as a polynomial of degree $M$ in the Hamiltonian parameter $g$, 
	\begin{equation}
		\tilde{\gamma}_k = \sum_{j=0}^M \zeta_{j, k} g^j, \quad \tilde{\beta}_k = \sum_{j=0}^M \eta_{j, k} g^j.
	\end{equation}
	Here, $\boldsymbol{\gamma}$ and $\boldsymbol{\beta}$ are indirectly optimized via the direct optimization of the vectors $\boldsymbol{\zeta}$ and $\boldsymbol{\eta}$. For the purposes of this study, we have selected $M=4$ as the degree of these polynomials. Leveraging the Fourier strategy alongside the polynomial representation enables the simultaneous global optimization of $\boldsymbol{\gamma}$ and $\boldsymbol{\beta}$ across various values of $g$. This approach facilitates the minimization of the energy function sum
	\begin{equation}
		\mathcal{E}_{\text{opt}} = \sum_{g \in \mathcal{G}_{\text{opt}}} \operatorname{Tr}\left( \mathcal{H}(g) \rho(\boldsymbol{\zeta}, \boldsymbol{\eta}; g)\right),
	\end{equation}
	where $\rho(\boldsymbol{\zeta}, \boldsymbol{\eta}; g) = U(\boldsymbol{\zeta}, \boldsymbol{\eta}; g) \rho_{\text{in}} U^\dagger(\boldsymbol{\zeta}, \boldsymbol{\eta}; g)$ are the optimized states. $\mathcal{E}_{\text{opt}}$ is theoretically lower bounded by the sum of the ground state energies on $\mathcal{G}_{\text{opt}}$, $\mathcal{E}_{\text{opt}} \geq \sum_{g \in \mathcal{G}_{\text{opt}}}E_{\text{gs}}(g)$. Starting with a variational quantum circuit with a single layer ($p=1$), we uniformly sample the initial values of $\boldsymbol{\zeta}, \boldsymbol{\eta}$ from a predefined range. The optimization of $\boldsymbol{\zeta}, \boldsymbol{\eta}$ is conducted through the Broyden–Fletcher–Goldfarb–Shanno algorithm~\cite{broyden1970the, fletcher1970an, goldfarb1970family, shanno1970conditioning}, and the outcomes are then used to guide subsequent optimizations for circuits with increasing depth $p+1$. With the increase of circuit depth $p$, the optimization deficit decreases, enhancing the precision of our estimates for the locations of critical points.

	Adopting a global optimization approach, in contrast to individual optimizations for each $g$ individually, confers significant benefits. Firstly, this strategy is more efficient, as, post-optimization, parameters corresponding to any given $g$ value can be readily generated, enabling the preparation of the associated state through circuit execution. Furthermore, global optimization ensures stability across closely related $g$ values, such as $g = 0.400$ and $g = 0.401$, allowing for the generation of states with consistent features. If the optimization were conducted separately for each $g$, even adjacent values of $g$ could potentially settle into different local traps, leading to distinct states. This divergence could result in highly unstable data unsuitable for machine learning analysis.

	The quantum optimization landscape is notoriously swamped with local traps~\cite{anschuetz2022quantum}, presenting significant challenges in finding the global minimum. In this work, the quantum states we obtain from the variational quantum optimization, denoted as $\rho(\boldsymbol{\zeta},\boldsymbol{\eta};g)$, are characterized as locally trapped states that have higher energy than the global minima for several reasons. Firstly, the Fourier strategy approach utilized, as discussed in Ref.~\cite{zhou2020quantum}, tends to be trapped in local minima when new parameters are extended by appending zero-vectors without integrating random perturbations upon the increase of $p$. Secondly, our approach is limited by the sampling of a single set of initial parameters for the optimization process, constraining the comprehensive exploration of the available parameter space. Lastly, we model the variables ${\Tilde{\gamma}_k}$ and ${\Tilde{\beta}_k}$ as low-degree polynomial functions of the Hamiltonian parameter $g$. This choice, while facilitating computational efficiency, limits full optimization for each $g$ value, thereby reducing the chances of finding the global minima.

	\begin{figure*}[t]
		\hrule
		\vspace{0.5em}
			\raggedright\textbf{Input:}  Optimization parameter range $[g_{\text{min}}, g_{\text{max}}]$ with sampling resolution $\delta g$, detection parameter range $[\hat{g}_{\text{min}}, \hat{g}_{\text{max}}]$ with sampling resolution $\delta \hat{g}$, window size $w$, and the set of Pauli operators $\mathcal{P}_f$ whose expectation values form the feature vector.
			\begin{algorithmic}[1] 
				\State Define optimization \& detection grid points $\mathcal{G}_{\text{opt}} = \{g_{\text{min}}, g_{\text{min}} + \delta g, \ldots, g_{\text{max}}\}$, $\mathcal{G}_{\text{det}} = \{\hat{g}_{\text{min}}, \hat{g}_{\text{min}} + \delta \hat{g}, \ldots, \hat{g}_{\text{max}}\}$. \Comment{Setup}
				\State Establish a variational quantum circuit with parameters as polynomial functions of $g$ with coefficients ${\boldsymbol{\zeta}, \boldsymbol{\eta}}$. Denote the output state as $\rho(\boldsymbol{\zeta}, \boldsymbol{\eta}; g)$. \Comment{Initialization}
				\State Optimize $\{\boldsymbol{\zeta}, \boldsymbol{\eta}\}$ to minimize $\mathcal{E}_{\text{opt}} = \sum_{g \in \mathcal{G}_{\text{opt}}} \operatorname{Tr}\left( \mathcal{H}(g) \rho(\boldsymbol{\zeta}, \boldsymbol{\eta}; g)\right)$. \Comment{Finding locally-trapped states}
				\For{each $g$ in $\mathcal{G}_{\text{det}}$}
				\State Prepare optimized quantum state $\rho(\boldsymbol{\zeta}, \boldsymbol{\eta}; g)$. 
				\State Measure Pauli terms in $\mathcal{P}_f$ and record the vector of expectation values $\boldsymbol{f}(g)$. \Comment{Record feature vectors for ML analysis}
				\EndFor
				\State Initialize training loss records $\mathcal{L} = \varnothing$. \Comment{Prepare to record training losses}
				\For{each $g$ in $\{\hat{g}_{\text{min}} + w \cdot \delta \hat{g}, \ldots, \hat{g}_{\text{max}} - w \cdot \delta \hat{g}\}$}
				\State Initialize training dataset $\mathcal{D} = \varnothing$. \Comment{Setup training set for each window}
				\For{each $\Tilde{g}$ in $\{g - w \cdot \delta \hat{g}, \ldots, g - \delta \hat{g}\}$}
				\State Assign label $-1$ to $\boldsymbol{f}(\Tilde{g})$, add $(\boldsymbol{f}(\Tilde{g}), -1)$ to $\mathcal{D}$. \Comment{Label pre-transition states}
				\EndFor
				\For{each $\Tilde{g}$ in $\{g + \delta \hat{g}, \ldots, g + w \cdot \delta \hat{g}\}$}
				\State Assign label 1 to $\boldsymbol{f}(\Tilde{g})$, add $(\boldsymbol{f}(\Tilde{g}), 1)$ to $\mathcal{D}$. \Comment{Label post-transition states}
				\EndFor
				\State Train supervised learning regressor on $\mathcal{D}$, classify phase based on $g$. \Comment{Use ML to identify phase transitions}
				\State Record, append training loss to $\mathcal{L}$. \Comment{Log training loss for analysis}
				\EndFor
				\State Analyse $\mathcal{L}$ for phase transitions, critical values. \Comment{Identify phase transitions}
			\end{algorithmic}
			\raggedright\textbf{Output:} Count and locations of detected phase transitions, and the corresponding order parameters.
		\vspace{0.5em}
		\hrule
		\caption{Hybrid Quantum-ML Algorithm for Phase Transition Detection.}
		\label{alg: Phase_Detection}
	\end{figure*}
	
	\vspace{1em}
	
	\paragraph{Classical loss landscape} The second stage of our methodology involves constructing a depiction of the classical machine learning loss function landscape across various magnitudes of the parameter $g$ by employing classical regressors. Our objective is to pinpoint critical points of phase transitions, which can be identified by analysing the valleys in this landscape. The number of these valleys is utilized as an approximation for the total count of phase transitions. To facilitate this exploration, we set up a detection grid, $\mathcal{G}_{\text{det}}$, consisting of a sequence of points $\{\hat{g}_{\text{min}}, \hat{g}_{\text{min}} + \delta \hat{g}, \hat{g}_{\text{min}} + 2\delta \hat{g}, \ldots, \hat{g}_{\text{max}}\}$ with $\delta \hat{g}$ being the detection sampling resolution. For each $g$ in $\mathcal{G}_{\text{det}}$, we prepare the trapped quantum state $\rho(\boldsymbol{\zeta},\boldsymbol{\eta};g)$, then measure and record the feature vector, $\boldsymbol{f}(g)$, which comprises the expectation values of selected Pauli operators, represented by the set $\mathcal{P}_f$, from this state.

    Our framework, termed \textit{presupposed phase label regression}, aligns philosophically with the learning by confusion approach~\cite{Nieuwenburg2017A, Arnold2022Replacing, Arnold2024Mapping} and effectively delineates phase boundaries using learnable order parameters. It identifies phase transition boundaries based on the performance fluctuations of the learning algorithm. The analysis employs a sliding window technique with a predetermined window size, $w \in \mathbb{N}^+$, to systematically explore the parameter space. Initially, we create an empty set to record training losses, denoted as $\mathcal{L} = \varnothing$. Subsequently, for each value of $g$ within the recalibrated range $[\hat{g}_{\text{min}} + w \cdot \delta \hat{g}, \hat{g}_{\text{max}} - w \cdot \delta \hat{g}]$, the process is as follows: we start with an empty training set $\mathcal{D} = \varnothing$; for each $\Tilde{g}$ in the range $\{g - w \cdot \delta \hat{g}, \ldots, g - \delta \hat{g}\}$, we assign a label of $-1$ and add the pair $(\boldsymbol{f}(\Tilde{g}), -1)$ to $\mathcal{D}$; similarly, for each $\Tilde{g}$ in $\{g + \delta \hat{g}, \ldots, g + w \cdot \delta \hat{g}\}$, we assign a label of 1 and incorporate the pair $(\boldsymbol{f}(\Tilde{g}), 1)$ into $\mathcal{D}$. Following this, we train a supervised learning regressor (such as LASSO or Transformer) using the dataset $\mathcal{D}$, record the training loss, and append this loss to $\mathcal{L}$. The presence of valleys within $\mathcal{L}$ indicates potential phase transitions, as they represent points where the supervised learning model discerns a significant distinction between pre-transition and post-transition states, effectively leveraging learnable order parameters. The selection of the window size $w$ is critical: if too small, the detection may lack stability; conversely, if too large, the detection may lack sensitivity and precision. The selection of $w$ is influenced by the complexity of our model, allowing for adaptive adjustments to secure an optimal window size. Notably, our methodology exhibits significant robustness to variations in hyperparameters, thereby demonstrating its stability across a broad range of computational settings. For more detailed analysis, refer to the SM. 
	
	The pseudocode of the algorithm is shown in Fig.~\ref{alg: Phase_Detection}. 
	
	\vspace{1em}
	
	\paragraph{Noise robustness} Quantum gate noise is an inherent challenge in current quantum devices~\cite{preskill2018quantum}, usually impacting the accuracy and reliability of quantum computations. Nevertheless, our framework exhibits substantial robustness to such disturbances, especially when the circuit noise scales linearly each feature vector. This is a practical assumption considering that local noise often approximates a global depolarizing noise channel as the circuit depth increases~\cite{deshpande2022tight, dalzell2024random}. The following theorem illustrates the robustness of our algorithm against such quasi-global depolarizing (Quasi-GD) noise under specific conditions:
	\begin{theorem}[Robustness to Quasi-GD Noise]\label{theorem:noise_robustness}
		Let $\boldsymbol{\zeta}$ and $\boldsymbol{\eta}$ be optimized parameters, with $\rho_{\text{in}}$ representing the density matrix of the input state. For a Hamiltonian parameter $g$, consider $\rho(\boldsymbol{\zeta}, \boldsymbol{\eta}; g)$ as the corresponding ideal output state, and $\mathcal{N}_{\boldsymbol{\zeta}, \boldsymbol{\eta}; g}(\cdot)$ as the corresponding noisy quantum channel. If for any $g$ in $\mathcal{G}_{\text{det}}$ and any Pauli operator $O$ in $\mathcal{P}_f$, the expectation values satisfy
		\begin{equation}
			\operatorname{Tr}\left(O \Lambda_{\epsilon}\left(\rho(\boldsymbol{\zeta}, \boldsymbol{\eta}; g)\right)\right) = \operatorname{Tr}\left(O \mathcal{N}_{\boldsymbol{\zeta}, \boldsymbol{\eta}; g}\left(\rho_{\text{in}}\right)\right),
		\end{equation}
		where $\Lambda_{\epsilon}(\cdot)$ denotes a global depolarizing channel with a fixed noise rate $\epsilon$, then the channel $\mathcal{N}_{\boldsymbol{\zeta}, \boldsymbol{\eta}; g}(\cdot)$ is defined as a quasi-global depolarizing channel with respect to $\{\rho(\boldsymbol{\zeta}, \boldsymbol{\eta}; g)\}$ and $\mathcal{P}_f$. Given these conditions, the machine learning algorithms—LASSO and Transformer—are capable of predicting critical points from noisy quantum data that are consistent with those predicted from ideal quantum data.
	\end{theorem}
	This theorem implies that for noise channels which act like global depolarizing channels with respect to the set of Pauli operators $\mathcal{P}_f$, the overall shape of the classical loss function remains invariant. In other words, the machine learning subroutine effectively performs automatic quantum error mitigation in detecting phase transitions. When local gate noise is modeled as depolarizing noise and the quantum circuit achieves sufficient depth, the local noise effectively transforms into global white noise~\cite{dalzell2024random}, the expectation values of the Pauli operators in $\mathcal{P}_f$ are uniformly attenuated, and the condition is satisfied. However, it is important to note that the set $\mathcal{P}_f$ of interest typically only comprises finite special operators, such as low-weight local Pauli terms. Consequently, the aforementioned condition for noise robustness might be satisfied with very shallow circuits.

    The proof of Theorem~\ref{theorem:noise_robustness} and the accompanying numerical investigations are detailed in the SM. Our findings indicate that in relatively shallow circuit regimes, the condition imposed by the theorem is approximately satisfied more effectively when the weight of the Pauli terms in $\mathcal{P}_f$ is small and when the circuit depth is reduced. This suggests that for practical implementations, one can achieve robust phase-transition detection without necessitating deep quantum circuits, provided that the feature set is appropriately chosen.

    Furthermore, it is crucial to emphasize that Theorem~\ref{theorem:noise_robustness} provides a sufficient but not a necessary condition for noise robustness. Our numerical experiments, presented in Section~\ref{Sec:1DTFIM} and the SM, also demonstrate that the Transformer-based machine learning model exhibits strong robustness against noise. Taking this one step further, even if the overall protocol is not inherently noise-robust, the shallow nature of our circuits ensures that quantum error mitigation techniques~\cite{Cai2023QuantumError}, such as zero-noise extrapolation and probabilistic error cancellation, can be efficiently applied to suppress the impact of noise. This provides additional flexibility and reliability in practical implementations.

	\subsection{Finite-Depth Extrapolation}
	\label{subsec:extrapolation}
	Finite-size scaling is a well-established and powerful technique to obtain precise estimates of the critical properties (such as critical exponents and critical points) of classical and quantum systems in the thermodynamic limit by inspecting how their properties vary as a function of the (finite) system size $n$~\cite{fisher1972scaling, newman1999montecarlo}.  This technique typically involves calculating size-specific observables—such as magnetization, susceptibility, or specific heat—and evaluating the resulting critical points $g_c(n)$ as functions of the system size $n$. From the scaling hypothesis, one can derive that the location of a critical point scales with the size of the system as
	\begin{equation}
		\Tilde{g}_c(n)=g_c(n\rightarrow \infty)+b n^{-\mu},
	\end{equation}
	enabling a predictive insight into phase transition critical points in the thermodynamic limit. Direct integration of finite-size scaling into our algorithm is achieved by executing the algorithm across various system sizes and employing polynomial fitting to forecast the phase transition critical value in the thermodynamic limit.
	
	Furthermore, in our work we introduce an extrapolation approach, termed finite-depth extrapolation. This method involves executing our algorithm across a spectrum of circuit depths $p$, within a specifically chosen system size $n$. For each $p$, the value of $n$ is chosen to ensure that quantum information propagation is localized without experiencing boundary effects~\cite{okada2023classically}. For instance, in one-dimensional systems with nearest-neighbour interactions where each layer promotes single-site quantum information propagation, we set $n \geq 2p + 2$ to prevent boundary effects. This approach allows the expectation values of certain geometrically central local observables to align with those anticipated for an infinitely large system. Therefore, this allows us to focus on the impact of circuit depth rather than system size. Similar finite-depth scaling techniques have been explored in ground-state preparation and the simulation of imaginary-time critical dynamics~\cite{bravoprieto2020scaling, zhang2024universal}.

	Through the application of machine learning techniques, an effective order parameter is discerned, and the landscape of the loss function provides an estimation of the phase transition critical point across different values of $p$. Interestingly, in specific scenarios it seems that the location of the critical point scales exponentially with $p$ rather than polynomially. This suggests that if a circuit with $n=\text{poly}(p)$ suffices to avoid boundary effects, then the location of the critical point converges to the thermodynamic limit value exponentially with $n$. Observation~\ref{statement:exponential_precision} provides more details, and this assertion is supported by numerical analyses in Sec.~\ref{Sec:1DTFIM}.
	\begin{observation}[Exponential Precision]\label{statement:exponential_precision}
		Consider an $n$-qubit quantum system with low-dimensional connectivity, structured such that a $p=\mathrm{poly}(n)$-layer circuit effectively avoids boundary effects. If there exists a local observable serving as an order parameter for a phase transition, the utilization of quantum optimization might allow the achievement of an estimate for the critical point with exponential accuracy, denoted as $\exp(n)$. This level of accuracy significantly exceeds the precision, $\mathrm{poly}(n)$, obtainable through direct preparation of the finite system's ground state.
	\end{observation}

	The intuition for the weak dependence on finite size effects is the following. In a gapped system, the correlation of operators decays exponentially with distance~\cite{hastings2006spectral}. Then it should be feasible to generate the same correlations in a system whose size is just on the order of the correlation length $\xi$. Away from a phase boundary, this correlation length is independent of the system size. Closer to a phase boundary, the correlation length scales on the order of $n$, so simulating a system with depth $p$ on the order of $n$ should suffice to minimize the finite size effects.	
	
	Application of the finite-depth extrapolation, as described by
	\begin{equation}
		\label{eq:depth_scaling}
		\Tilde{g}_c(p)=g_c(p\rightarrow\infty)+ c e^{-\nu p},
	\end{equation}
	allows for the estimation of the critical point for an infinitely deep circuit, that is, the phase transition critical point in the thermodynamic limit, $g_c(p\rightarrow\infty)$. Notably, as we discussed above, $g_c(p\rightarrow\infty)$ coincides with the result for the ground state in an infinitely large system, denoted $g_c(n\rightarrow\infty)$. Consequently, $g_c$ is used henceforth to denote both the thermodynamic and infinite circuit depth critical points.

	The rationale for employing exponential fitting in our finite-depth extrapolation is underpinned by two significant observations. Firstly, as identified in Ref.~\cite{dreyer2021quantum}, the energy density of the output states from variational quantum optimization exhibits an exponential dependence with respect to the circuit depth $p$. This characteristic facilitates the determination of critical points, which can be discerned as the derivative of the energy with respect to the Hamiltonian parameter. Secondly, as detailed in Sec.~\ref{Sec:1DTFIM}, our numerical tests reveal that the infidelity between the states produced by quantum optimization and the true ground states near the infinite-size critical point diminishes exponentially with increasing $p$. Consequently, the discrepancies in the local order parameters are also expected to reduce exponentially with increasing $p$. Therefore, the finite-depth extrapolation technique we utilize is based on this exponential decay in discrepancies:
	\begin{equation}
		\label{eq:exponential_extrapolation}
		|\Tilde{g}_c(p)-g_c| \sim 1/\text{poly}(2^p),
	\end{equation}
	thus exhibiting exponential convergence toward the infinite-size critical point as the depth increases.

\subsection*{Acknowledgements}
The authors would like to thank Jan Lukas Bosse, Brian Flynn, and other members of the Phasecraft team, as well as Marco Cerezo, Michael Meth and Philipp Schindler for their helpful discussions. This project has received funding from the European Research Council (ERC) under the European Union’s Horizon 2020 research and innovation programme (grant agreement No.\ 817581) and from InnovateUK grant 44167.

\bibliography{bibliography}

\onecolumngrid
\newpage 

\begin{center}
	\textbf{\large Supplementary Information for ``Unveiling quantum phase transitions from traps in variational quantum algorithms"}
\end{center}
\setcounter{equation}{0}
\setcounter{figure}{0}
\setcounter{table}{0}
\setcounter{page}{1}
\makeatletter
\renewcommand{\thetable}{S\arabic{table}}
\renewcommand{\thefigure}{S\arabic{figure}}
\renewcommand{\theequation}{S\arabic{equation}}

\twocolumngrid

\section{Robustness of Variational Parameter Optimization to Noise}\label{Sec: Noisy Optimization}
In this section, we present a detailed numerical study of the impact of noise on the optimization of variational parameters in the 1D transverse-field Ising model (TFIM). To simulate realistic hardware conditions, each $R_{ZZ}$ rotation gate in the variational circuit is immediately followed by two corresponding local depolarizing noise channels modeled by
\begin{equation}
    \mathcal{N}_{\mathrm{DP}}(\rho)=\left(1-\epsilon^{(\mathrm{DP})}\right) \rho+\epsilon^{(\mathrm{DP})} \frac{I}{4}
\end{equation}
with $\epsilon^{(\text{DP})}=0.01$ acting on corresponding qubits.  We fix a shallow circuit with depth $p=2$ and perform the optimization under two scenarios: one using an ideal (noiseless) circuit, and the other using the noisy circuit described above. In both cases, we compare the optimized parameters and the resulting quantum states to assess whether the presence of noise significantly alters the optimized variational parameters.

\begin{figure}[bh]
	\centering
	\includegraphics[width=7.4cm]{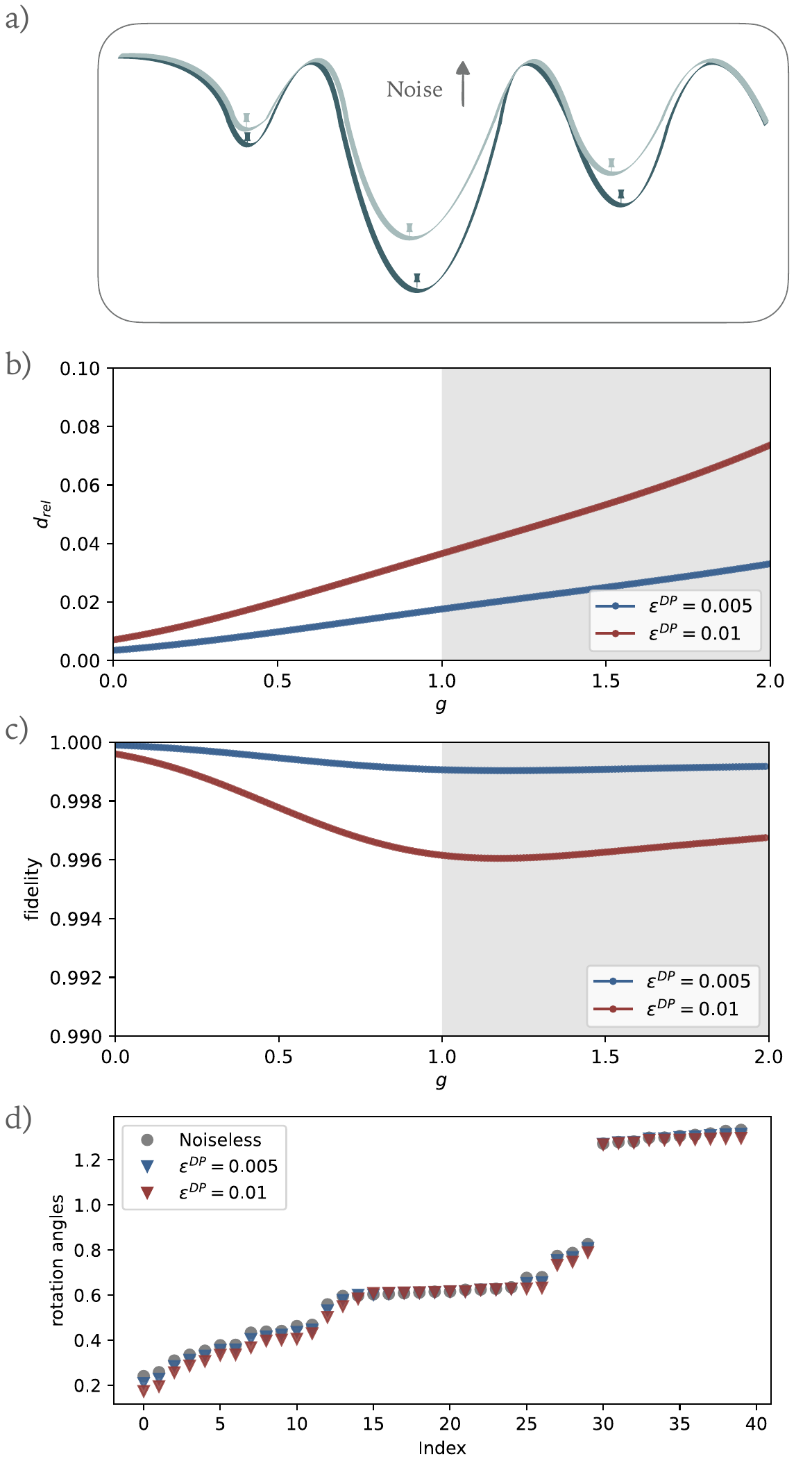}
	\caption{Robustness of optimized variational parameters in the 1D TFIM under local depolarizing noise. (a) A schematic illustration showing that, as noise increases, the energy landscape becomes flatter, while the positions of the local minima remain largely unaffected. (b) The relative distance $d_{\text{rel}}$ between the ideally optimized rotation angles and those obtained from noisy optimization is plotted as a function of the transverse field $g$ for two noise levels, $\epsilon^{(\text{DP})}=0.005$ (blue markers) and $\epsilon^{(\text{DP})}=0.01$ (red markers). (c) The state fidelity, computed by embedding the optimized parameters from both noisy and noiseless circuits into an ideal $14$-qubit circuit, is plotted versus $g$ for the same two noise levels. Fidelity values are consistently above $0.996$, indicating minimal deviation. (d) A representative comparison of 40 sample sets of rotation angles from noisy and noiseless optimization.}
	\label{fig:Noisy Optimization}
\end{figure}

Fig.~\ref{fig:Noisy Optimization}(a) presents a schematic that illustrates our expectation: as noise increases, the energy landscape becomes flatter, but the local minima remain nearly unchanged. Panels (b)–(d) then provide numerical results.  In panel (b), we plot the relative distance between the ideally optimized rotation angles and those obtained from noisy optimization as a function of the transverse field $g$ for two noise rates, $\epsilon^{(\text{DP})}=0.005$ and $\epsilon^{(\text{DP})}=0.01$. The relative distance is defined as
\begin{equation}
d_{\text {rel}}=\frac{\sqrt{\| \boldsymbol{\gamma}_ \text { noisy }-\boldsymbol{\gamma}_\text {ideal }\left\|_2^2+\right\| \boldsymbol{\beta}_\text { noisy }-\boldsymbol{\beta}_{\text {ideal }} \|_2^2}}{\sqrt{\| \boldsymbol{\gamma}_\text { ideal }\left\|_2^2+\right\| \boldsymbol{\beta}_\text { ideal } \|_2^2}},
\end{equation}
where $\left\|\cdot\right\| _2$ is the Euclidean norm,
which remains below 0.074 across all values of $g$. Panel (b) shows that when the optimized parameters (obtained from the noisy circuit) are embedded into an otherwise noiseless circuit with $14$ qubits, the fidelity between the states produced by the noisy and ideal parameters remains high (all above $0.996$) for both noise levels. In panel (c), we present detailed comparisons of 40 samples of rotation angles, demonstrating that the optimized angles obtained from noisy and noiseless circuits are in close agreement.

These results illustrate that moderate levels of local depolarizing noise have a minimal impact on the optimized variational parameters, thereby preserving the ability of our algorithm to accurately identify the effective order parameter and the critical point. Ref.~\cite{liu2023stochastic} has shown that the inherent stochasticity of noise may help the quantum optimization process avoid strict saddle points. Although our study does not directly exploit this potential benefit, our findings of robust performance under moderate noise levels suggest that some amount of noise might even facilitate finding better local minima. We leave a detailed investigation of these effects for future work.

\begin{figure}[tbh]
	\centering
	\includegraphics[width=7.5cm]{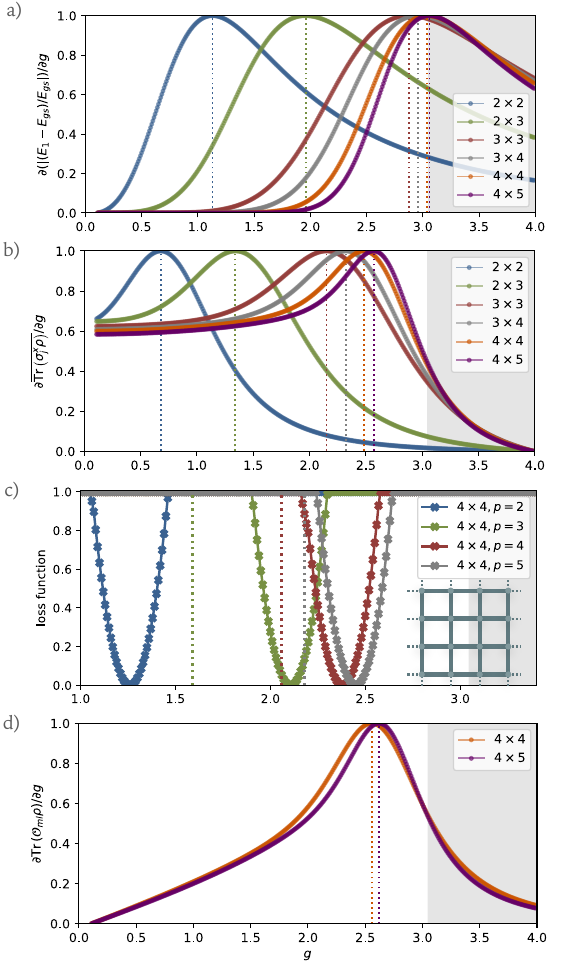}
	\caption{(a) Derivatives of the relative spectral gap as a function of the transverse field $g$ for system sizes $2 \times 2$, $2 \times 3$, $3 \times 3$, $3 \times 4$, $4 \times 4$, and $4 \times 5$ under periodic boundary conditions. (b) Derivatives of the $x$-magnetization as a function of $g$ for the same systems. (c) LASSO loss landscapes obtained from trapped states on a $4 \times 4$ grid with periodic boundary conditions for circuit depths $p=2, 3, 4$, and $5$. Dashed lines denote the critical boundaries estimated via $x$-magnetization. (d)  Derivatives of the learned order parameter as a function of $g$ for the ground state data of $4 \times 4$ and $4 \times 5$ grids. Dashed lines in panels (a), (b), and (d), denote estimated critical boundaries.}
	\label{fig:TFIM-Appendix}
\end{figure}

\section{Larger System Investigations for the 2D TFIM}\label{Sec: TFIM-44}
In this section, we present additional numerical results for larger qubit grids with periodic boundary conditions. We consider several lattice configurations, including $2 \times 2$, $2 \times 3$, $3 \times 3$, $3 \times 4$, $4 \times 4$, and $4 \times 5$ systems. and analyse the estimation of the phase transition critical point using different metrics.

\begin{figure*}[ht!]
	\begin{lstlisting}[label=lst:set_transformer, basicstyle=\ttfamily\small\color{gray}, keywordstyle=\color{gray}]
		import torch.nn as nn
	\end{lstlisting}
	\begin{lstlisting}[ basicstyle=\ttfamily\small\color{black}, keywordstyle=\color{black}]
		
		class SAB(nn.Module):
		def __init__(self, d_model, num_heads):
		super(SAB, self).__init__()
		self.attention = nn.MultiheadAttention(d_model, num_heads)
		self.norm = nn.LayerNorm(d_model)
		self.dropout = nn.Dropout(0.1)
		
		def forward(self, X):
		X_norm = self.norm(X)
		attention_output, _ = self.attention(X_norm, X_norm, X_norm)
		attention_output = self.dropout(attention_output)
		output = self.norm(X + attention_output)
		return output
	\end{lstlisting}
	
	\begin{lstlisting}[label=lst:set_transformer2, basicstyle=\ttfamily\small\color{gray}, keywordstyle=\color{gray}]
		class SetTransformerRegressor(nn.Module):
		def __init__(self, d_model, num_heads, num_layers, fc_intermediate_dim=16):
		super(SetTransformerRegressor, self).__init__()
		self.sab_blocks = nn.ModuleList([SAB(d_model, num_heads) for _ in range(num_layers)])
		self.fc1 = nn.Linear(d_model, fc_intermediate_dim)
		self.fc2 = nn.Linear(fc_intermediate_dim, 1)
		self.activation = nn.ReLU()
		
		def forward(self, src):
		src = src.unsqueeze(1) 
		for sab in self.sab_blocks:
		src = sab(src)
		src = src.squeeze(1)
		src = self.activation(self.fc1(src))
		output = self.fc2(src).squeeze(-1)
		return output
	\end{lstlisting}
	\caption{Exemplary code for the core architecture of the Set Transformer, utilized for detecting quantum phase transitions in the eSSH model. This includes implementations of both the Self-Attention Block (SAB) and the overall regression model structure.}
	\label{fig:code_transformer}
\end{figure*}

In panel (a) of Fig.~\ref{fig:TFIM-Appendix}, we plot the derivatives of the absolute relative spectral gap $(E_1-E_{\text{gs}})/|E_{\text{gs}}|$ as a function of the transverse field $g$ for the aforementioned system sizes. The estimated critical points from these derivatives are approximately 1.13, 1.96, 2.88, 2.96, 3.04, and 3.06 for the $2 \times 2$, $2 \times 3$, $3 \times 3$, $3 \times 4$, $4 \times 4$, and $4 \times 5$ systems, respectively. Notably, the last two estimates are very close to the thermodynamic limit value, $g_c = 3.04438$~\cite{blote2002cluster, schmitt2022quantum}. Panel (b) shows the derivatives of the $x$-magnetization as a function of $g$ for the same set of system sizes. The corresponding estimated critical points are 0.68, 1.34, 2.15, 2.32, 2.48, and 2.57. These estimates are clearly less accurate than those obtained from the relative spectral gap, highlighting the limitations of local order parameters in finite systems.

We further implement our LASSO-based algorithm on a $4 \times 4$ grid with periodic boundary conditions for circuit depths $p=2, 3, 4$, and $5$. The average fidelities for these depths are 0.5473, 0.7336, 0.8674, and 0.9293, respectively. In panel (c) of Fig.~\ref{fig:TFIM-Appendix}, we display the LASSO loss landscapes for each circuit depth. The estimated critical points from the loss landscapes are 1.26, 2.11, 2.38, and 2.45 for $p=2, 3, 4$, and $5$, respectively. For comparison, we also include, using dashed lines, the critical boundaries obtained from the $x$-magnetization of the trapped states. Notably, for $p=2$, the $x$-magnetization fails to provide an estimate, and for $p=3,4,5$, the estimates derived from $x$-magnetization are considerably worse than those obtained via our algorithm. The dominant order parameter identified by our method in this case is $\mathcal{O}_{\mathrm{ml}}=\sum_{i,j}\sigma_{i,j}^x \sigma_{i+2,j+1}^x$.

Finally, in panel (d), we present the derivative of the learned order parameter (i.e., $\mathcal{O}_{\mathrm{ml}}$) with respect to $g$, computed from the ground state data for both the $4 \times 4$ and $4 \times 5$ grids. The derivative curves obtained using the learned order parameter yield more accurate estimates of the critical point (2.56 and 2.62, respectively) compared to those based on $x$-magnetization.

\section{Exemplary Set Transformer Code and Complexity Analysis}\label{Sec: Set Transformer Code}
In Fig.~\ref{fig:code_transformer} we provide the exemplary Python code for the Set Transformer used in Section~\ref{Sec:eSSH}. Central to this architecture is the Self-Attention Block (SAB), which processes the input data through the mechanism of self-attention. In this setup, the normalized input, $X\_{\text{norm}}$, serves simultaneously as the Query, Key, and Value. This configuration enables each element of the input feature interact with every other feature, facilitating a comprehensive internal representation that captures the underlying relationships.

\begin{figure}[tbh]
	\centering
	\includegraphics[width=8.5cm]{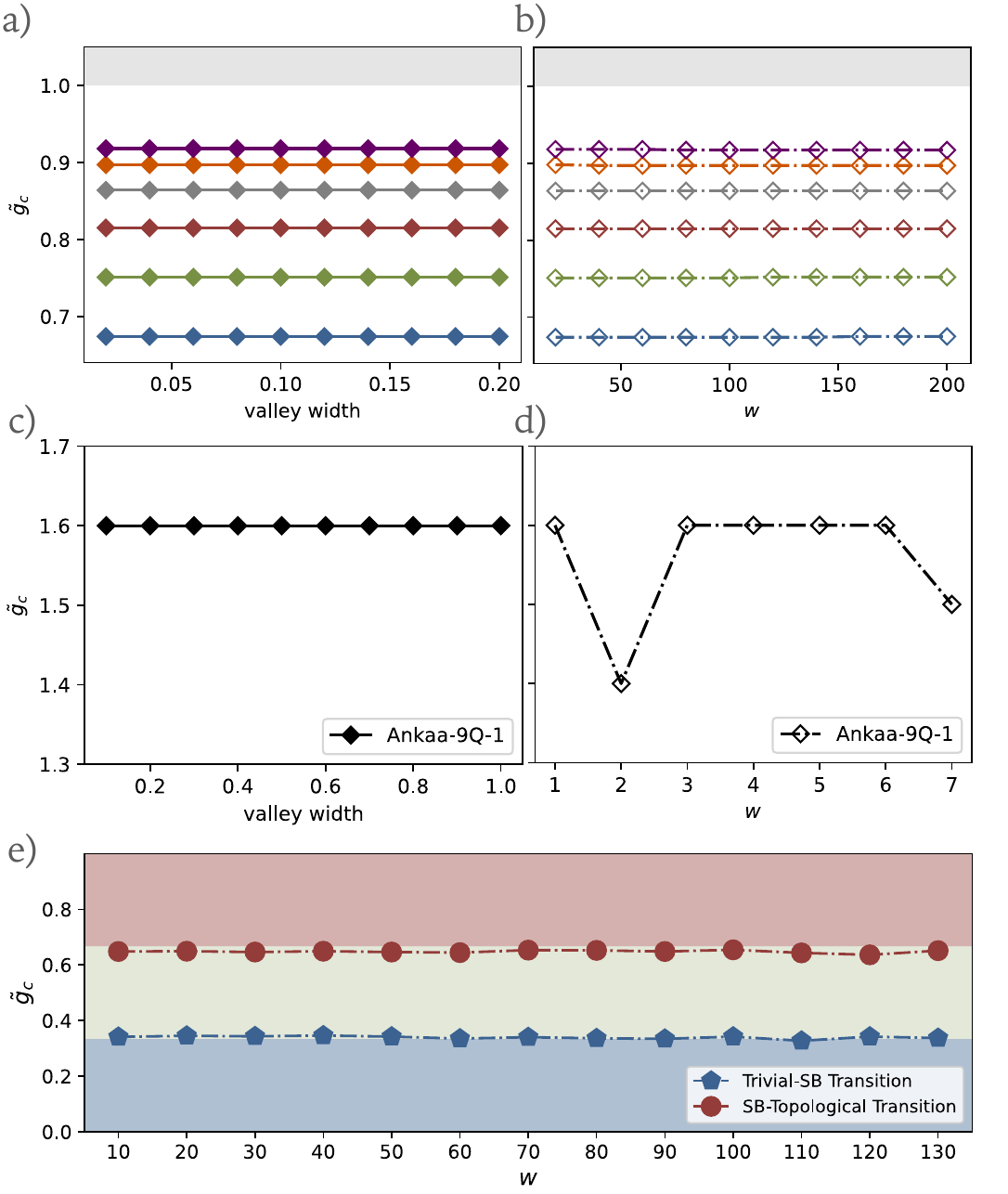}
	\caption{Predicted critical points for varying valley widths and window sizes, demonstrating the algorithm's robustness with respect to hyperparameters $\lambda$ and $w$ across diverse settings. (a) and (b) focus on processing numerical data for the 1D transverse-field Ising model (TFIM) using LASSO. In both panels, colours correspond to the different values of $p$ defined in Fig.~\ref{fig:TFIM}(a, b). (c) and (d) present results from processing experimental data for the 2D TFIM using LASSO. (e) showcases the Set Transformer algorithm applied to the extended Su-Schrieffer-Heeger (eSSH) model with $\delta = 4$, highlighting its response to changes in $w$.}
	\label{fig:hyperparameters}
\end{figure}

In our proposed framework, a set of optimization grid points, whose total number we denote by \(N\), is chosen as potential candidates for detecting phase transitions. For each valid center \(g\), the algorithm collects a local dataset of size \(2w\), formed by \(w\) parameter values below \(g\) (labeled \(-1\)) and \(w\) parameter values above \(g\) (labeled \(+1\)), and then trains a supervised model on these samples. Let \(\ell = \ell(n)\) be the total number of observables (Pauli operators or otherwise) whose expectation values are measured and fed into the model. In practice, \(\ell\) may be chosen to grow polynomially in the system size \(n\) if only local or few-body operators are included, or it may grow exponentially when higher-body correlations are considered.

In each training session, we fit a classical regression model—LASSO or a Transformer—to a labeled dataset of size \(2w\).   For LASSO, we employ a coordinate descent solver, which incurs a per-iteration computational cost of order \(w\ell^2\). If the algorithm requires \(s\) iterations to converge, the cost of a single LASSO training run is \(\mathcal{O}(s\, w\, \ell^2)\). When \(s\) and \(w\) remain nearly constant or grow only slowly with \(n\), \(\ell(n)^2\) becomes the dominating factor.

For Transformers inspired by the Set Transformer framework, the forward/backward pass of a single attention layer scales as \(\mathcal{O}(\ell^2\,d)\), where \(d\) is the embedding dimension. In the presence of \(h\) attention heads, the overall cost for one layer is approximately \(\mathcal{O}(h\,\ell^2\,d)\). If \(\tau\) is the number of epochs or gradient updates, each Transformer training session has an overall cost of \(\mathcal{O}(\tau\,h\,\ell^2\,d)\), possibly with additional constants from feedforward blocks. Because the dataset size is \(2w\), \(\tau\) generally stays modest  in most practical settings, so again \(\ell(n)^2\) often drives the scaling.

Since the sliding-window procedure trains a separate model at each center \(g\), these costs must be multiplied by the number of centers \(N_c\). If boundary effects in the parameter grid are negligible, \(N_c\) is of the same order as \(N\). Consequently, applying LASSO at all window centers yields a total cost of \(\mathcal{O}\bigl(N_c \times s\, w\, \ell^2\bigr)\), while a Transformer-based approach costs \(\mathcal{O}\bigl(N_c \times \tau\,h\,\ell^2\,d\bigr)\). In typical setups, \(s\), \(\tau\), \(h\), \(d\), and \(w\) are constants or grow only mildly with \(n\), making \(\ell(n)^2\) and \(N_c\) the primary contributors to the runtime. If \(\ell(n)\) can be chosen to scale polynomially with \(n\) (or remain approximately constant), and \(N_c\) is a constant independent of \(n\), then the overall complexity of the sliding-window detection algorithm scales polynomially with \(n\). Under these assumptions, the protocol remains computationally feasible for moderate system sizes relevant to current or near-term quantum simulations.

Finally, we empirically evaluated the runtime of our LASSO-based detection procedure on the 1D TFIM for system sizes \(n=4,6,\ldots,16\). As shown in Table~\ref{tab:lasso_runtime}, we fixed the window size at \(w=30\) and the number of sampled observables at \(\ell=30\), and scanned \(N_c=1940\) boundary points. Despite increasing \(n\), the total runtime remains nearly constant at around 1--1.4 seconds, indicating that once \(w\) and \(\ell\) are fixed, our post-processing overhead does not appreciably grow with the system size.

\begin{table}[H]
\centering
\renewcommand{\arraystretch}{1.2}
\setlength{\tabcolsep}{3pt}
\begin{tabular}{c|ccccccc}
\hline
System size & 4 & 6 & 8 & 10 & 12 & 14 & 16 \\ \hline
Runtime (s) & 1.381 & 1.395 & 1.308 & 1.405 & 1.381 & 1.296 & 1.405 \\ \hline
\end{tabular}
\caption{Total runtime (in seconds) for the LASSO-based detection of phase boundaries in the 1D TFIM, with \(N_c=1940\), \(\ell=30\), and \(w=30\) kept fixed, across various system sizes \(n\).}
\label{tab:lasso_runtime}
\end{table}

From a theoretical standpoint, the iteration count or epoch number in these ``black box" ML models depends largely on the complexity of distinguishing phases in feature space rather than simply on the system size. In many physically relevant models, this complexity does not necessarily scale with \(n\). Consequently, even in regimes where a full classical simulation of the quantum state becomes intractable, the ML-based post-processing is expected remain tractable, provided it only requires distinguishing phases based on a polynomial number of measured features.

\section{Robustness of Hyperparameter Sensitivity}\label{Sec: Robustness}
This section evaluates the stability and robustness of our hybrid quantum optimization-machine learning algorithm against variations in two key hyperparameters: the regularization parameter $\lambda$ within the LASSO algorithm, which inversely influences the width of the loss function landscape's valleys, and the window size $w$, crucial for both the LASSO and Transformer-based algorithms for setting the range of phase labels $-1$ and $1$. It is essential to avoid excessively high values of $\lambda$, as this would result in all coefficients $\boldsymbol{\kappa}$ shrinking to zero, thereby reducing the LASSO's target cost function to a constant value of 1. We analyse the algorithm's performance using three illustrative examples: LASSO for the numerical data from the 1D TFIM, LASSO for the experimental data from the 2D TFIM, and the Set Transformer for the eSSH model with $\delta = 4$, where two phase transitions are observed.

The results, displayed in Fig.~\ref{fig:hyperparameters}, include panels (a) and (b) for the 1D TFIM, panels (c) and (d) for the 2D TFIM, and panel (e) for the eSSH model. Panel (a) shows the predicted critical points for various valley widths with a constant window size of $w=30$, demonstrating small sensitivity to this hyperparameter in LASSO as all lines remain constant. Panel (b) illustrates the predicted critical points across different window sizes with a set valley width of 0.06, where the maximum deviation observed is only $0.01$. Panel (c) displays results for different valley widths at a window size of $w=4$, consistently converging to $1.6$. Panel (d) examines various window sizes at a valley width of $5$, where the maximum variation is $0.2$, but consistently approximates to the most likely value of $1.6$. Panel (e) evaluates the Transformer's efficacy on the eSSH model with $\delta = 4$, considering various window sizes. It shows that the standard deviations associated with predictions of phase transitions from trivial to symmetry-broken (SB) phases and from SB to topological phases are 0.00525 and 0.00482, respectively, both indicating low variability.

These findings underline the limited sensitivity of our algorithm to changes in $\lambda$ and $w$, confirming its robustness across different computational environments for both numerical and experimental data.

\begin{figure*}[tbh]
	\centering
	\includegraphics[width=16cm]{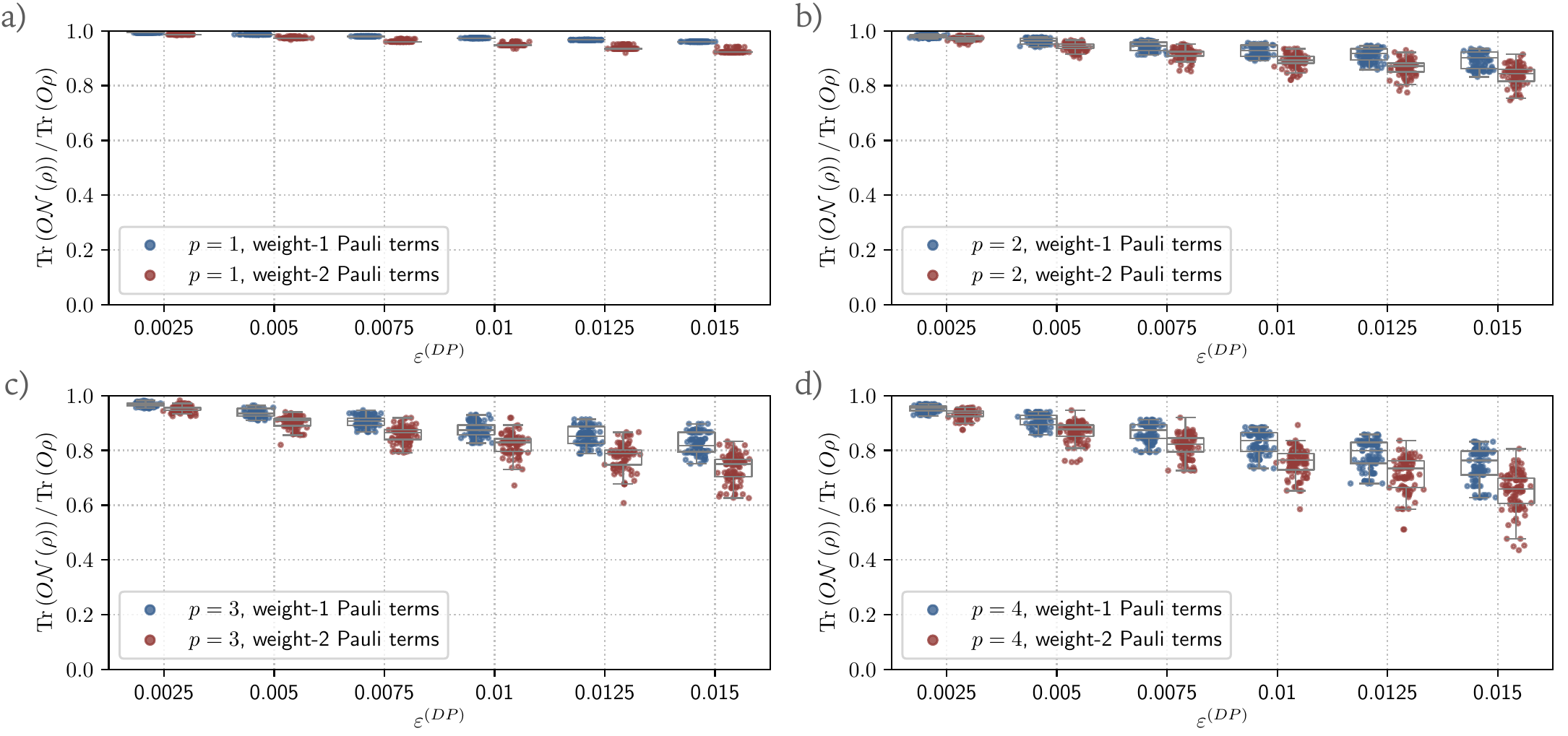}
	\caption{Effect of local depolarizing noise on Pauli expectation ratios in Hamiltonian variational ansatz circuits for the 1D TFIM. The four panels correspond to a different circuit depth: (a) $p=1$, (b) $p=2$, (c) $p=3$, and (d) $p=4$.  For each panel, the horizontal axis represents the local depolarizing noise rate $\epsilon^{(\text{DP})}$ and the vertical axis displays the ratio \(R(O_j, \rho)= \operatorname{Tr}\left(O_j \mathcal{N}_{\text{DP}}(\rho)\right)/\operatorname{Tr}\left(O_j \rho\right)\) for a randomly chosen Pauli operator $O_j$. Each data point is associated to the state $\rho$ corresponding to one of 200 quantum circuits with a randomly sampled set of rotation angles $\{\boldsymbol{\beta}, \boldsymbol{\gamma}\}$. Blue points correspond to measurements of weight-1 Pauli operators, while red points correspond to weight-2 Pauli operators. The results demonstrate that for shallower circuits and lower noise rates, the variance of the ratio is minimal, validating the quasi-global depolarizing noise approximation for the randomly chosen observables.}
	\label{fig:QGD}
\end{figure*}

\section{Robustness to Quasi-Global Depolarizing Noise}\label{sec:proof_noise_robustness}
\begin{proof_theorem}{\ref{theorem:noise_robustness}}
	The feature vector $\boldsymbol{f}(g)$, input into the machine learning model, consists of the expectation values of Pauli operators from the set $\mathcal{P}_f$:
	\begin{equation}
		\boldsymbol{f}(g)=\left(\operatorname{Tr}\left(O_1 \rho(\boldsymbol{\zeta}, \boldsymbol{\eta} ; g)\right),\ldots,\operatorname{Tr}\left(O_\ell \rho(\boldsymbol{\zeta}, \boldsymbol{\eta} ; g)\right)\right),
	\end{equation}
	where each $O_j$ is a Pauli operator from the set $\mathcal{P}_f$.
	
	Assuming the noise behaves like a global depolarizing channel $\Lambda_\epsilon$, defined by:
	\begin{equation}
		\Lambda_\epsilon(\rho) = (1-\epsilon) \rho + \epsilon \frac{I}{2^n},
	\end{equation}
	the expectation value of a Pauli operator $O$ under this noise model becomes:
	\begin{equation}
		\operatorname{Tr}\left(O \Lambda_\epsilon(\rho(\boldsymbol{\zeta}, \boldsymbol{\eta}; g))\right) = (1-\epsilon) \operatorname{Tr}(O \rho(\boldsymbol{\zeta}, \boldsymbol{\eta}; g)) + \frac{\epsilon}{2^n} \operatorname{Tr}\left(O\right).
	\end{equation}
	Since the trace of non-identity Pauli operators  is zero, this simplifies to:
	\begin{equation}
		\operatorname{Tr}\left(O \Lambda_\epsilon(\rho(\boldsymbol{\zeta}, \boldsymbol{\eta}; g))\right) = (1-\epsilon) \operatorname{Tr}(O \rho(\boldsymbol{\zeta}, \boldsymbol{\eta}; g)).
	\end{equation}
	Consequently, the noisy feature vector $\boldsymbol{f}_{\text{noisy}}(g)$ is a scaled version of the original feature vector:
	\begin{equation}
		\boldsymbol{f}_{\text{noisy}}(g) = (1-\epsilon) \boldsymbol{f}(g).
	\end{equation}

	When the noisy feature vector $\boldsymbol{f}_{\text{noisy}}(g)$ is processed using LASSO, we can adjust the regularization coefficient $\lambda$ by multiplying it by $(1-\epsilon)$. Considering the LASSO cost function as shown in Eq.~\eqref{eq:Lasso_loss}:
	\begin{equation}
		\mathcal{C}(\boldsymbol{\kappa}, \lambda) = \left(\frac{1}{4 w} \sum_{i=1}^{2 w}\left(l_i-\kappa_0-\sum_{j=1}^{\ell} \kappa_j f_{i j}\right)^2+\lambda \sum_{j=1}^{\ell}\left|\kappa_j\right|\right)
	\end{equation}
	for the coefficients $\boldsymbol{\kappa} = \left(\kappa_0, \kappa_1, \kappa_2, \dots, \kappa_\ell \right)$, the same cost function $\mathcal{C}$ can be achieved with the input $\boldsymbol{f}(g)$ and the adjusted coefficients:
	\begin{equation}
		\boldsymbol{\kappa}_{\text{noisy}} = \left(\kappa_0,  \frac{\kappa_1}{1-\epsilon}, \frac{\kappa_2}{1-\epsilon}, \dots, \frac{\kappa_\ell}{1-\epsilon}\right).
	\end{equation}
	Thus, with adequate optimization, the classical loss landscape and the predicted critical points remain unchanged as long as we scale the regularization parameter $\lambda$ by multiplying it with the factor $(1-\epsilon)$.
	
	When processing the noisy feature vector $\boldsymbol{f}_{\text{noisy}}(g)$ through the Transformer model, the initial step involves normalization of feature vectors. This step rescales data and effectively mitigating the scaling introduced by the noise, ensuring that the learning model's output remains unaffected by the scaling effect of the quasi-global depolarizing noise.
\end{proof_theorem}

To empirically validate the conditions specified in Theorem~\ref{theorem:noise_robustness}, we conducted numerical simulations using the Hamiltonian variational ansatz designed for the one-dimensional transverse-field Ising model (1D TFIM). In our simulation framework, each two-qubit rotation gate is subject to local depolarizing noise with a rate of $\epsilon^{(\text{DP})}$,  which results in an effective two-qubit error rate of approximately $2\epsilon^{(\text{DP})}$.

We generated a representative ensemble of circuit outputs by randomly sampling 200 distinct sets of rotation angles $\{\boldsymbol{\beta}, \boldsymbol{\gamma}\}$. For each resultant quantum state $\rho$, we randomly select one Pauli operator $O_j$ from the set of Pauli operators with weight 1 or 2. To ensure that only significant contributions are considered, we imposed the condition \(|\operatorname{Tr}(O_j \rho)| > 0.2\). This thresholding procedure serves to filter out Pauli terms with negligible expectation values, which would otherwise be disproportionately affected by small noise-induced fluctuations.

The key metric we analyse is the ratio of the expectation value of the noisy state to that of the ideal (noiseless) state, defined as $R(O_j, \rho)=\operatorname{Tr}\left(O_j \mathcal{N}(\rho)\right)/\operatorname{Tr}\left(O_j \rho\right)$.
We evaluated this ratio for a system size of $n=10$ qubits and varying circuit depths $p = 1, 2, 3, 4$, while systematically adjusting the gate noise rate to different values $\epsilon^{(\text{DP})} = 0.0025, 0.005, 0.0075, 0.01, 0.0125, 0.015$. The results, depicted in Fig.~\ref{fig:QGD}, display the distribution of the computed ratios for different configurations. Each subplot corresponds to a specific circuit depth, with blue markers representing ratios for weight-1 Pauli operators and red markers representing ratios for weight-2 Pauli operators.

We observe that the variance of the computed ratios is smaller when: (i) the circuit depth is low, (ii) the noise rate is small, and (iii) the observables are low-weight. The third behavior can be understood theoretically by noting that, for shallow circuits, the noise affecting individual qubits can be treated as statistically independent. For an operator that acts nontrivially on $k$ qubits, the overall noise-induced attenuation is approximately given by the product $(1-\epsilon')^k$ for certain $\epsilon'$. For low-weight operators (small $k$), the expectation values are only marginally reduced and exhibit low variance. In contrast, for high-weight operators, the multiplicative noise effects compound, leading to significantly lower expectation values and higher relative variance. As the circuit depth increases, the variance in the ratios grows, indicating that the quasi-global depolarizing approximation underlying Theorem~\ref{theorem:noise_robustness} becomes progressively less valid. These results imply that by carefully selecting circuit depths and favoring lower-weight Pauli operators for $\mathcal{P}_f$, the impact of noise can be more effectively mitigated in practical implementations.

\end{document}